\documentclass[a4paper,twocolumn,english,aps,showpacs,letter]{revtex4}
\newcommand{\bu}{\mathbf{u}}
\newcommand{\ba}{\hat{\mathbf{a}}}
\newcommand{\bW}{\mathbf{W}}

\newcommand{\hH}{\hat{H}}
\newcommand{\hR}{\hat{R}}
\newcommand{\hQ}{\hat{Q}}
\newcommand{\hU}{\hat{U}}
\newcommand{\hX}{\hat{X}}

\newcommand{\ha}{\hat{a}}
\newcommand{\hb}{\hat{b}}
\newcommand{\hr}{\hat{r}}

\usepackage{amsmath}
\usepackage{graphicx,color}
\begin{document}

\title[RH method]{Fermi gas response to  time-dependent perturbations}

\author{N. d'Ambrumenil$^{1,2}$ and B. Muzykantskii$^1$}

\affiliation{$^1$Department of Physics, University of Warwick, Coventry CV4 7AL,
 UK \\
$^2$School of Physics and Astronomy, University of Birmingham, Birmingham B15 2TT, UK 
}

\date{\today{}}

\begin{abstract}
We describe the Riemann-Hilbert (RH) approach to
computing the long-time response of a Fermi
gas to a time-dependent perturbation. The approach
maps the problem onto a non-commuting
RH problem. The method is non-perturbative, quite 
general and 
can be used to compute the Fermi
gas response in driven (out of equilibrium) as
well as equilibrium systems. It has the 
appealing feature of working directly with 
scattering amplitudes defined at the Fermi surface rather
than with the bare Hamiltonian.
We illustrate the power of the method by rederiving standard
results for the core-hole and open-line Greens functions
for the equilibrium Fermi edge singularity (FES) problem. We then
show that the case of the non-separable potential can
be solved non-perturbatively with no more effort than
for the separable case. We compute the corresponding results
for a biased (non-equilibrium) model tunneling device, 
similar to those used in single
photon detectors, in which a photon absorption
process can significantly change the conductance of the barrier.
For times much larger than the inverse bias across the device, the response 
of the Fermi gases in the two electrodes shows that the
equilibrium  Fermi
edge singularity is smoothed, shifted in frequency and
becomes polarity-dependent.
These results have a simple interpretation
in terms of known results for the equilibrium case but with 
(in general complex-valued) combinations of elements of the 
scattering matrix  replacing the equilibrium phase shifts.
We also consider the shot noise spectrum of a tunnel junction
subject to a time-dependent bias and demonstrate that the 
calculation is essentially the same as for the FES problem.
For the case of a periodically driven device
we show that the noise spectrum for the 
Coherent States of Alternating Current found in \cite{IvaLL97} 
can be easily obtained using this approach.
\end{abstract}
\maketitle

\section{Introduction}
An approach to the study of 
the quantum statistics of an {\bf arbitrary} single-particle observable
in a  Fermi gas has been recently described in~\cite{MA03}. 
We refer to it  as the RH approach, as it
reduces the calculation of a determinant describing the  quantum statistics 
of an observable to the solution of an (in general) non-Abelian 
Riemann-Hilbert (RH) problem. 
The relation between
such determinants and RH problems has been known
for a long time, and
has been used extensively in studies of quantum
inverse scattering problems \cite{Korepin_QISM93}.
As a result, a lot is known about the non-abelian
RH problem  \cite{DeiftZhou},
and much of this can be taken over directly to the study of the 
quantum statistics of Fermi gases.

The RH approach gives an expression for the 
distribution  function of an
observable in a Fermi gas perturbed by a 
time-dependent potential. 
To illustrate the method, one of us used
it to prove a long-standing conjecture, first stated
in  \cite{LevitovJETP93}, that the two sources of
shot noise in a biased point contact, namely
fluctuations in the number of attempts to tunnel
through the barrier and fluctuations in the number of reflections, are
statistically independent \cite{MA03}. 
We have also used the method to study how
non-equilibrium effects alter the Fermi Edge Singularity 
in a tunnel junction \cite{MdAB03}.

The response of a Fermi gas to a 
time-dependent perturbing potential
is a central problem in condensed
matter physics. It has been tackled in many
different contexts often with different approaches.
For systems out of equilibrium, such as quantum
pumps, perturbative approaches,
based on the Keldysh formalism, have been used, while
for systems in equilibrium it has been possible to find
exact solutions in some limiting cases by solving the
equations of motion directly \cite{Mahan67,ND69,Schotte+Schotte69,YY82}.
One of the advantages of the RH method is that it applies
equally to all such problems and
therefore offers the prospect of a unified approach 
to computing the time-dependent response of all 
observables in Fermi gases.

Setting up the description of a problem in the RH framework 
is quite straightforward. Given the solution of the
single-particle scattering problem, 
the response of the
Fermi gas reduces to the computation of a determinant
of an operator taken over single-particle states 
occupied in the initial configuration. 
(The generalization of the method to the more general case 
in which the initial state is given in terms of a density matrix 
rather than a single quantum state should be possible
but has not yet been formulated.) The evaluation of this
determinant then reduces to the solution
of a Riemann Hilbert problem.
The solution  is in general a matrix-valued 
function analytic everywhere  except across a cut, along which
the function is discontinuous. The discontinuity 
is fixed by the  driving force or perturbation acting 
upon the system \cite{MA03}.
From the point of view of the Keldysh formalism the method
performs a non-trivial re-summation of all relevant diagrams with the help
of the solution of the corresponding RH problem.
In the abelian case, when the discontinuity
function commutes with itself at all points along the cut,
the solution is given in terms of an integral. 
The classic solution of the FES problem
\cite{ND69,Schotte+Schotte69} is the
simplest example of this solution.
In the non-abelian case, the solution to the RH problem is not 
known in general, although asymptotic solutions exist.
These are valid for response frequencies small compared to those
present in the discontinuity function.

Here we explain the RH approach in some detail. To illustrate
the power of the method we start by showing how the solution of
the equilibrium FES problem \cite{ND69,Schotte+Schotte69,CN71}
is derived. We 
then show the generalization of this problem to include the
case where the `impurity' potential mixes scattering states of the 
unperturbed problem---the case of a non-separable potential---and
deal explicitly with the case when the impurity potential
gives rise to a bound state. This problem
was treated initially in \cite{YY82,Matveev-Larkin}, in
a calculation which
solved directly the Dyson-like equation 
for the appropriate Green's functions. 
In the RH formulation of this problem 
the discontinuity function,
although matrix-valued, is constant and commutes with itself.
As a consequence, the solution to the RH problem is trivial to derive
and yields the standard results of \cite{YY82,Matveev-Larkin}
with no more work than for the case separable potential case.
We show how these results are changed in a non-equilibrium
situation. In both the equilibrium and non-equilibrium
cases, we compute both the core-hole Green's function
reported in \cite{MdAB03} and the open line contribution.
Finally we show how the states which minimize the shot noise
in a periodically driven quantum pump---the so-called
Coherent States of Alternating Current (CSAC's)
\cite{IvaLL97}---can be described using the RH method.

\section{Perturbing the Fermi Gas}
\begin{figure}[tbp]
  \centering
\includegraphics[width=8cm]{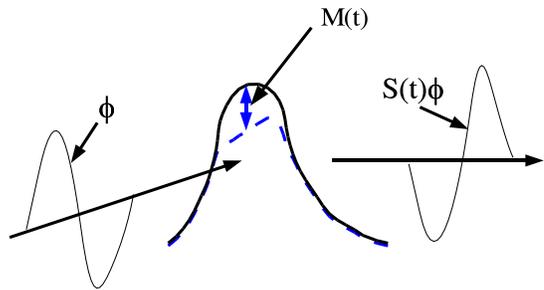}
  \caption[]{Schematic illustration of the generic scattering
problem. Particles impinge on a localized potential which
is independent of time for times $t<t_0$ and $t>t_f$.
For times $t_0<t<t_f$ the potential varies as a function of time.
The instantaneous value of the potential at time $t$ is shown
as the dashed  curve and the difference measured with respect
to the time-independent potential is denoted by
$M(t)$. When the potential varies slowly (condition
\ref{eq:slowness_condition}), the effect of the scattering
potential on an incident partial wave (shown schematically
as $\phi$) is simply multiplication by the scattering
matrix corresponding to the instantaneous value of $M(t)$.
 }
  \label{fig:scattering}
\end{figure}

We consider a system in which particles
impinge upon a localized potential (see Fig \ref{fig:scattering}). 
The potential
is time-independent for all times $t<t_0$ and
$t>t_f$. For times $t_0 < t < t_f$ the potential varies.
We take our basis to be the eigen states of the system
with the localised potential at its  value.
The states are labeled by their single-particle energy $\epsilon$
($\hbar=1$) and a channel index $i$. We will consider 
the corresponding
annihilation operator, $a_{i\epsilon}$, as the $i$'th component
of the vector $\ba_\epsilon$.
The Hamiltonian of the system is then
\begin{eqnarray}
\hH(t) & = & \hH_0
 + \sum_{\epsilon,\epsilon'} \ba^+_\epsilon 
M(t,\epsilon,\epsilon') \ba_{\epsilon'} 
\nonumber \\
\hH_0 & = & \sum_\epsilon \epsilon \ba^+_\epsilon \ba_\epsilon. 
\label{eq:H(t)}
\end{eqnarray}
Here $M(t,\epsilon,\epsilon')=0$ when $t<0$ or $t>t_f$.
(In the following, for any operator $\hat{O}$, we will denote by $O$
the matrix of $\hat{O}$ taken between the single-particle basis states.)

We will be interested in the total effect of the perturbation,
{\it ie\/} what is the final state of the system for
$t > t_f$ given the initial state at $t=0$.
This requires a knowledge of the effect on the initial
many-body state of the time-development
operator $\hU (t_f)$, where
\begin{equation}
i \frac{d\hU}{dt} = \hH(t) \hU (t), \quad \hU (0)=1.
\label{eq:dU/dt}
\end{equation}
Because the Hamiltonian $\hH(t)$ in (\ref{eq:H(t)}) is quadratic,
the effect of $\hU (t_f)$ is fully characterized by its effect
on the set of {\bf single-particle} scattering states,
$a_{i\epsilon'}^+ |\rangle$:
\begin{equation}
  \hU (t_f) \ba_{\epsilon'}^+ |\rangle = \sum_{\epsilon} 
e^{-i\epsilon t_f} \sigma(\epsilon,\epsilon') \ba_{\epsilon}^+ |\rangle,
  \label{eq:motion}
\end{equation}
where $\sigma(\epsilon,\epsilon')$ is some unitary 
$N \times N$ and $|\rangle$ is the true vacuum with
no particles in the system \cite{Note_on_sigma}.

When computing the response of the Fermi gas to the time-dependent
potential, we will need to compute expectation values
of the type:
\begin{equation}
\chi_R = \langle 0 |\hR|0\rangle.
\label{eq:measurement}
\end{equation}
Here  $|0 \rangle$
is the state of the Fermi gas before the perturbation is applied
and $\hR$ is an operator (or operator product) related to an 
obvervable of interest.
In general,
the $\hR$ in (\ref{eq:measurement}) 
will involve the time-development operator
$\hU(t_f)$. 
For example, in the case of the shot noise spectrum
of a tunneling barrier \cite{MA03}, the interest is in 
the statistics of the charged transferred from one
electrode to the other. If $\hQ_1$ is the charge in
the first electrode  then
the expectation value of
$\hR= \hU^\dagger (t_f) e^{i\lambda \hQ_1} \hU (t_f) e^{-i\lambda \hQ_1}$,
yields the generating function for moments of the distribution
of charge transferred out of channel 1 (into channel 2) during
the period between $t=0$ and $t=t_f$. 
In the case of the FES problem \cite{MdAB03}, the core-hole Green's function
(see below) is related to the overlap 
$\langle 0| \hU_0^\dagger(t_f) \hU (t_f)|0\rangle $ where $\hU_0(t_f)$ is
the time-development operator for $\hH_0$ in (\ref{eq:H(t)}).
This overlap is an expectation value of the type (\ref{eq:measurement})
with $\hR = \hU_0^\dagger(t_f) \hU (t_f)$.

The effect of $\hU (t_f)$ acting on the single-particle states of
the basis is given by the unitary matrix $\sigma(\epsilon,\epsilon')$
defined in (\ref{eq:motion}). The matrix $\sigma(\epsilon,\epsilon')$
can be related to the scattering matrix, $S(t,E)$, for a particle 
with energy $E$
evaluated on the {\it instantaneous\/} value of the potential  $M(t)$
in (\ref{eq:H(t)}). This reflects the fact that $S(t,E)$
encodes all the information in the potential variations $M(t)$.
This  relation will be complicated in general.
However, when
\begin{equation}
    \hbar \frac{\partial S^{-1} }{\partial t} \frac{\partial S}{\partial E} 
\ll 1,
\label{eq:slowness_condition}
\end{equation}
the relation between $\sigma$ and $S$ becomes simple:
\begin{eqnarray}
\sigma_{l\epsilon l'\epsilon'} & = & S_{l \epsilon l' \epsilon'} 
\nonumber \\
S_{l \epsilon l' \epsilon'} & = & \frac{1}{2 \pi \sqrt{\nu_l \nu_{l'}}} 
\int dt S_{ll'}(t,E) 
e^{i(\epsilon-\epsilon')t},
\label{eq:FT(S)}
\end{eqnarray}
where $E=(\epsilon+\epsilon')/2$ and
$\nu_i$ is the density of states in channel $i$. The result (\ref{eq:FT(S)})
shows that the total scattering amplitude from state $k$ in
channel $n$ to $k'$ in channel $n'$ is just the Fourier
transform of the scattering matrix $S(t)$ evaluated
on the instantaneous value of the potential.
This result is well-known. It was used implicitly 
to solve the  FES problem in the presence of a separable
potential \cite{CN71} and is often
used to simplify studies of quantum pumps. 

A brief derivation of the condition (\ref{eq:slowness_condition}) 
is given in \cite{MA03}. The condition
can be understood heuristically as follows
(see also \cite{IvaLL97}). 
We consider the incoming wave-packet to be a partial
wave in channel $n$ of the basis,
in which $S$ is diagonal before the perturbation is
switched on. After impinging on the 
potential, the partial waves scattered from channel
$n$ into channel $n'$ will take
a time of order of the corresponding Wigner delay time  to 
pass out of the region where the potential acts. 
The condition (\ref{eq:slowness_condition}) is equivalent to
the requirement that the scattering matrix does not change 
significantly during this delay time. If this condition is
satisfied, the relation (\ref{eq:FT(S)}) also has a simple
interpretation. A wave with energy $\epsilon'=k'v$ 
in channel $n'$ incident at time $t$ 
on the potential (which is 
assumed to be localized around the origin)
will have amplitude at the origin proportional to $e^{-ik'vt}$ 
where $v$ is the velocity of the incident wave.
Waves will emanate from the source at the 
origin with amplitude  $S_{nn'}(t)e^{-ik'vt}$ in channel
$n$. If dispersion effects are small this will lead 
to a waveform  $S_{nn'}(t-(r/v'))e^{-ik'v'(t-(x/v'))}$. Decomposing
this into waves of the form $e^{ik(x-v't)}$ as 
$t \rightarrow t_\infty$ gives the result (\ref{eq:FT(S)}).
The normalization factor $1/\sqrt{\nu_l \nu_l'}$ 
is included so that in the case where the
incoming flux ($\sim v|\psi_{n}|^2$)
is totally  scattered 
into channel $n'$ the scattering amplitude is 1.

\subsection{Fermi Sea at $T=0$}
The calculation of the response of the Fermi gas
to the time-dependent perturbation reduces 
to the computation of the expectation
value $\chi_R=\langle 0 |\hR|0\rangle$ in (\ref{eq:measurement}).
In the following we will assume that $|0 \rangle$
is a single Slater determinant. In this case 
the computation of  $\chi_R$ requires
the evaluation of a single determinant:
\begin{equation}
\langle 0 |\hR|0\rangle = \mbox{det}^{\prime} \mid R \mid ,
\label{eq:det}
\end{equation}
where the elements of $R$ are given by:
\begin{equation}
R_{ii'}(\epsilon,\epsilon') = 
\langle | \ha_{i\epsilon} \hR \ha^\dagger_{i'\epsilon'} |\rangle .
\label{eq:R_elements}
\end{equation}
and where the prime on the det indicates that the determinant
is taken only over states occupied in $|0\rangle$. 

When the initial Slater determinant
$|0\rangle$ corresponds to a filled Fermi sea,  
it is useful to introduce the Fermi distribution
as an operator with elements:
\begin{equation}
f_{\epsilon,\epsilon'} = \delta_{\epsilon,\epsilon'}\delta_{ii'} 
\theta(-(\epsilon-\mu)) .
\label{eq:Fermi_distn}
\end{equation}
Here $\mu$ is the chemical potential, which we take to be zero.
(For the non-equilibrium
problems discussed later the chemical potential 
can vary according to the channel index. However,
a time-dependent change of basis replaces the differing
chemical potentials in the different electrodes by an
additional time-dependence in the scattering matrix $S(t)$,
so that there is no loss of generality in assuming $\mu=0$.)
In a block notation
that separates the states with positive and negative energies 
$f$ and $R$ become:
\begin{equation}
f = \begin{pmatrix}
  1 & 0 \\
  0 & 0
  \end{pmatrix} \qquad
  R = 
  \begin{pmatrix}
     R_{11} &  R_{12} \\
     R_{21} &  R_{22} 
  \end{pmatrix}.
\label{eq:blockF}
\end{equation}
and
\begin{equation}
\label{eq:block2}
 1-  f +  f  R = \begin{pmatrix}
 R_{11} &  R_{12} \\
0 &  1  
\end{pmatrix}.
\end{equation}
It then follows that 
\begin{equation}
\chi_R = \mbox{det}^{\prime} \mid R_{11} \mid
= \mbox{det} | 1-  f +  f  R |
\label{eq:full_det}
\end{equation}
where $\mbox{det}$ is now a determinant
taken over all states in the basis.

Expressing $\chi_R$ as the determinant of
$ 1-  f +  f  R$ taken over all states in the basis, allows us to write
\begin{equation}
\log{\chi_R}  =  \mbox{\rm Tr}\left( f \ln R  \right) +
\mbox{\rm Tr}\left(
\ln{( 1-  f +  f  R)} - f \ln {R } \right)
\label{eq:log_chi}
\end{equation}
In (\ref{eq:log_chi}), we have added and subtracted the term
$\log{\chi_R^{(1)}} \equiv 
\mbox{\rm Tr}\left( f \ln R  \right)$. This term consists 
of the diagonal elements of $\ln R $ summed 
over all occupied states 
in $|0\rangle$ and gives the contributions linear in $t_f$. 
It often has a simple physical interpretation.
In the FES
problem it yields the threshold shift (or change in the
ground state energy of the Fermi gas after the core hole
is created), while in the shot noise spectrum of the tunneling
barrier it can be shown to be related to the average transfer of 
charge across the barrier (the  Brouwer formula 
\cite{Brouwer98,MA03}).

The second term in (\ref{eq:log_chi}),
\begin{equation}
\log{\chi_R^{(2)}} \equiv  \mbox{\rm Tr}\left(
\ln{( 1-  f +  f  R)} - f \ln R \right), 
 \label{eq:log_chi_R^2} \\
\end{equation}
accounts for all the non-trivial effects associated with excitations close 
to the Fermi surface induced by the perturbation.  
(States far from the Fermi energy, when 
$f=1$ or $f=0$, make no contribution to this term. As a result
there are no problems associated with effects of the band edge 
or short time cutoff when computing this term.) 
In the case of the FES problem it describes the 
line-shape, while in the shot noise spectrum it gives all the
higher moments of the charge transfer distribution. 
When computing this term, we will later switch to the 
time-representation in which
\begin{equation}
R_{l\epsilon l'\epsilon'}  =  \frac{1}{2\pi \sqrt{\nu_l \nu_{l'}} } 
\int dt R_{ll'}(t,E)  e^{i(\epsilon-\epsilon')t},
\label{eq:R(t)}
\end{equation}
with (as in \ref{eq:FT(S)}) $E=(\epsilon+\epsilon')/2$. 
$R$ will normally involve  $\sigma$
or some simple combination of
$\sigma$ with itself and its inverse.
Provided the condition 
(\ref{eq:slowness_condition}) is satisfied,
we will be able to evaluate $\sigma$ by ignoring
the dependence of $S(t,E)$ on $E$.
(This $E$ dependence is not important
as states far from the Fermi energy do not contribute
to $\log{\chi_R^{(2)}}$.)
As a result the term
$\log{\chi_R^{(2)}}$ will depend only on the time-dependence
of the scattering matrix evaluated at the Fermi energy,
which we will denote by $S(t)$.

\subsection{The Riemann-Hilbert Problem}
Computing the second term in (\ref{eq:log_chi}), $\log{\chi_R^{(2)}}$,
is  the central task
in the evaluation of the response of the Fermi gas.
The non-trivial part of this  is finding the inverse of
$( 1-  f +  f  R)$ which can then be 
used in an integral representation for its logarithm.
This inverse can be written in terms of the
solution of an $N \times N$ matrix
Riemann-Hilbert problem, where $N$ is the length of the 
vector $\ba_\epsilon$, {\it ie\/} 
the number of channels in the problem (see \ref{eq:H(t)}).

A standard procedure for representing the logarithm of an infinite matrix,
such as the one on the right hand side of (\ref{eq:log_chi_R^2}),
introduces a $\lambda$-dependence for $R$ 
via \cite{lambda} 
\begin{equation}
R(\lambda) = \exp{( \lambda \log{R} )} 
\label{eq:R(lambda)}
\end{equation}
and then uses an integral over $\lambda$ to represent the logarithm:
\begin{equation}
\log{\chi_R^{(2)}} = \int_0^1 d\lambda 
  \mbox{\rm Tr}\left[ \left(
 ( 1-  f +  f  R)^{-1} f - f  R^{-1} 
 \right) \frac{dR}{d\lambda} \right] .
\label{eq:chi2}
\end{equation}
(The $\lambda$-dependence of $R$ introduced in (\ref{eq:R(lambda)})
is assumed in (\ref{eq:chi2}) although not written
explicitly.)

To compute the trace in (\ref{eq:chi2}), we switch to
a time representation in which a quantity $A$ becomes:
\begin{equation}
A_{ll'}(t,t') = \frac{1}{2\pi \sqrt{\nu_l \nu_{l'}}}
\int_{-\infty}^\infty \nu_l d\epsilon \int_{-\infty}^\infty \nu_{l'} d\epsilon'
A_{ll'}(\epsilon,\epsilon') 
\label{eq:time-representation}
\end{equation}
Now, the Fermi distribution (\ref{eq:Fermi_distn}) is no longer diagonal:
\begin{equation}
f_{ll'}(t,t') = \frac{i}{2\pi} \frac{\delta_{ll'}} {t-t'+i0} . 
\label{eq:f(t,t')}
\end{equation}
However, as we can neglect the $E$ dependence of $R(\lambda,t,E)$ and
$S(t,E)$ (see discussion after \ref{eq:R(t)}), 
$R$ and $S$ are now diagonal in $t$ and equal
to $R(\lambda,t,0)\delta(t-t')$ and $S(t,0)\delta(t-t')$ respectively.
In the time-representation, the product of two quantities 
requires matrix multiplication in the space of scattering channels 
together with an integral over the intermediate time
coordinate. Where
one of the quantities in the product
is diagonal in $t$ (for example $S$),
the integral over the intermediate time coordinate is of course 
trivial and the product reduces to the simple matrix
multiplication in the channel space.
The Tr now becomes a trace over
the scattering channels, which we denote by tr, and an integral over
the time coordinate, so that 
  \begin{equation}
\log{\chi_R^{(2)}}   =   \int_0^1 d\lambda 
  \int dt \mbox{\rm tr} \, \left[ \left(
 (1-  f + f  R)^{-1}  f - f  R^{-1} 
 \right) \frac{dR}{d\lambda} \right] .
\label{eq:time-trace}
\end{equation}
Here, when computing the diagonal (equal time) elements of 
$O$, one should take
$
\lim_{t\rightarrow t'} O(t,t').
$
If $A$ and $B$  are
diagonal in the time representation, it follows that
\begin{eqnarray}
\int dt \,  \mbox{\rm tr}
\left[ A, f \right] B
 & = &  \int dt \, \lim_{t'\rightarrow t} \,   \mbox{\rm tr}
\frac{i}{2\pi}
\left[ \frac{A(t)-A(t')}{t-t'+i0}  \right] B(t') \nonumber \\
& = & \frac{i}{2\pi} \int dt \,  \mbox{\rm tr} \,
 \frac{dA(t)}{dt} B(t) ,
\label{eq:dA/dt}
\end{eqnarray}
which is a result we use later.

The quantity $(1- f + f R)^{-1}$ 
can be written
in terms of the function $Y(t)$, which is a matrix in 
the channel space and which solves
an auxiliary Riemann-Hilbert problem. $Y(t)$ should be analytic
everywhere in the complex $t-$plane except on the interval
$[0,t_f]$ on the real axis along which  it satisfies
\begin{equation}
Y_-(t)Y_+^{-1}(t) = 
R(t) \qquad \mbox{where} \qquad Y_\pm = Y(t \pm i0). 
\label{eq:jump_condition} 
\end{equation}
In addition $Y$ should satisfy
\begin{equation}
Y \rightarrow 1 \qquad \mbox{when} \qquad |t|\rightarrow \infty. 
\label{eq:Y->1}
\end{equation}
These analytic properties together with (\ref{eq:f(t,t')}) yield
the useful identities
\begin{eqnarray}
f Y_- f & = & f Y_- \nonumber \\
f Y_+ f & = & Y_+ f 
\label{eq:fYf_relations}
\end{eqnarray}
Using these relations (and assuming that
$Y^{-1}$ is also analytic everywhere except
along the cut), it is then easy to verify that
\begin{equation}
( 1-  f +  f  R)^{-1} = Y_+\left( 
(1-f)Y_+^{-1} + fY_-^{-1} \right).
\label{eq:inverse_(1-f+fR)}
\end{equation}
As an aside, we note that $( 1-  f +  f  R)^{-1}$
is the solution to a singular integral equation,
with $f$ playing the role of the singular kernel of the Cauchy
type. It is well known that such integral equations
can be  solved using Carleman's method, which writes the
solution in terms of an analytic function satisfying a
Riemann-Hilbert problem  \cite{Muskhelishvili}. In
the one-channel the corresponding singular integral
equation for the case when $R(t)$ is constant between
$t=0$ and $t=t_f$ is the problem solved
in \cite{ND69,Schotte+Schotte69} when describing
the equilibrium FES.

Inserting (\ref{eq:inverse_(1-f+fR)}) into 
(\ref{eq:time-trace})
and using (\ref{eq:dA/dt}) we obtain 
\begin{equation}
\log{\chi_R^{(2)}} = \frac{i}{2\pi}  \int_0^1 d\lambda
  \int dt\, \mbox{\rm tr}\left\{ 
\frac{dY_+}{dt}Y_+^{-1}R^{-1} \frac{dR}{d\lambda} \right\} .
\label{eq:rhtime-trace}
\end{equation}
The integral over $t$ is over all times. However, as
$dR/d\lambda$ normally vanishes for $t>t_f$ and $t<0$, 
one often only needs to integrate from
$0$ to $t_f$.  

Equations (\ref{eq:rhtime-trace}) and (\ref{eq:log_chi})
map the characterisation of the response $\chi_R$ 
in (\ref{eq:measurement}) onto an integral involving the solution,
$Y(t)$, of a Riemann-Hilbert (RH) problem
(\ref{eq:jump_condition} and \ref{eq:Y->1}). An appealing
feature of this formulation is that these formulas
apply for any choice of variable $\chi_R$ provided that 
the Fermi gas (or gases) is initially in its ground
state and apply for many non-equilibrium cases as well.
If $R(t)$ commutes with itself at different 
values of $t$, the solution of the 
RH problem can be written in closed form. 
Although there is no solution
for the general case, a lot is known about such 
non-commuting problems including some asymptotic solutions 
valid when $t_f^{-1}$ is much smaller than any characteristic
frequency in $R(t)$ \cite{DeiftZhou}.

\section{Fermi Edge Singularity}
In this section we show how all the known results for
the equilibrium Fermi Edge Singularity (FES) 
follow directly from the formula (\ref{eq:rhtime-trace}).
Within our formalism
the case of non-separable channels considered
in \cite{YY82} and again in \cite{Matveev-Larkin} 
is no more complicated than the separable case. We 
will then discuss how these results are changed in the
nonequilibrium case.

\subsection{Equilibrium}
The FES problem was first considered in the context
of the X-ray absorption spectrum of a metal \cite{Mahan67}. When
a photon  creates a core hole in a metal,
the Fermi gas is affected by the potential of the core hole
leading to the excitation of particle-hole pairs.
The absorption line expected in the absence of the Fermi gas
becomes a threshold with a singularity in the absorption
spectrum as a function of $\omega - \omega_0 > 0$:
\begin{equation}
I(\omega) \sim \mid \omega - \omega_0 \mid^{-\alpha} ,
\label{eq:FES}
\end{equation}
where $\omega_0$ is the threshold frequency
for absorption. It turns out that similar
singularities are seen in
the distribution of energy absorbed by the Fermi gas in
response to any rapid change in potential and not just
in x-ray absorption experiments. For example, the consequences
of the FES are also seen in a tunnel junction. 
As the energy absorbed by the Fermi gas when switching, 
is an important characteristic of the device,
establishing how the FES changes in such tunneling devices
is important for understanding
fluctuations in energy transfer across
such devices.
The FES is also thought to be 
related to  the apparent absence of detailed balance
in Random Telegraph Signals \cite{Cobden-Muzykantskii95}.

The Hamiltonian for the photon absorption experiment 
is \cite{ND69}
\begin{equation}
\hH = \epsilon_0 \hb^\dagger \hb +
\sum_\epsilon \epsilon \ba^+_\epsilon \ba_\epsilon
+ \sum_{\epsilon' \epsilon} \ba^\dagger_\epsilon 
V(\epsilon,\epsilon') \ba_{\epsilon'} \, \hb\hb^\dagger+ \hH_X
\end{equation}
with the operators $\ba$ as  in (\ref{eq:H(t)}) and
$V(\epsilon,\epsilon')$ an $N\times N$ matrix. The operator
$\hb^\dagger$ is the creation operator corresponding to  the core
state and the coupling to
the X-ray field is described semiclassically by
\begin{eqnarray}
\hH_X & = & \sum_\epsilon \bW_\epsilon \cdot \ba^\dagger_\epsilon 
\hb e^{i\omega t} 
+ h.c. \nonumber \\
    & \equiv & \hX   e^{i\omega t}
\label{eq:xray_field}
\end{eqnarray} 
The absorption spectrum is proportional to the real
part of the Fourier transform of the response function 
\begin{equation}
S(t_f) = \langle0|T\{ \hX(t_f) \hX(0) \} |0\rangle ,
\label{eq:S(t)}
\end{equation}
with $T$ the time-ordering operator.
$S(t_f)$ can be computed from
the  core-hole Green's function \cite{ND69}
\begin{equation}
G(t_f)  =  \langle 0 | T\{ \hb^\dagger(t_f) \hb(0) \}| 0 \rangle
\label{eq:core-hole} 
\end{equation}
and the function
\begin{equation}
F(t_f)  =  \sum_{\epsilon,\epsilon'} 
 \langle 0 | T\{ \hb^\dagger(t_f)
 (\bW_{\epsilon'}^* \cdot \ba_{\epsilon'}(t_f))  \}
 (\bW_\epsilon \cdot\ba^\dagger_\epsilon(0)) \hb(0) | 0 \rangle .
\label{eq:open-line}
\end{equation}
Coventionally a minus sign is included in the definition of
$F$ and $G$. However, as we will only deal with the absorption
case here and take $t_f > 0$, it is easier to work from
these definition. We have also left out the conventional
factor of $i$ in the definitions of these Green's functions
as in  \cite{ND69}.

The calculation
of $F$ and $G$ reduces to a one-body scattering problem
\cite{ND69,Friedel69}. As far
as the Fermi gas is concerned 
the  role of the core hole is to switch on the scattering
potential  $V(\epsilon,\epsilon')$ at time $0$ and switch it off 
again at $t_f$.
As such, the problem is clearly in the form of (\ref{eq:H(t)}) with
$M(t,\epsilon,\epsilon') = 0$ for $t>t_f$ and $t<0$ and 
$M(t,\epsilon,\epsilon') = V(\epsilon,\epsilon')$ for 
$0<t<t_f$. The corresponding scattering matrix $S(t,E)$
switches between the identity when the core-hole is
absent and some constant value $S^e(E)$ when it is present. The
asymptotic behavior of the response 
at large $t_f$ (when $\omega-\omega_0 \ll \xi_0^{-1}$)
is determined by states with energies 
close to the Fermi surface. For these states we 
assume that the variation of $S^e(E)$ with $E$ can be neglected
so that the condition (\ref{eq:slowness_condition}) is
satisfied. (The limit $\xi_0 t_f \gg 1$ is the one
considered in \cite{ND69}.) 

The calculation of $G$ is one of the simplest calculations
within the RH approach. $G$ is the expectation value of the operator 
$\hR$ in (\ref{eq:measurement}) with:
\begin{equation}
\hR = \hU_0^\dagger(t_f) \hU (t_f).
\label{eq:R_G}
\end{equation}
As the matrix elements of $\hU (t_f)$ are just
$e^{-i\epsilon t_f} \sigma_{\epsilon \epsilon'}$,
it follows from (\ref{eq:det}) that \cite{CN71}  
\begin{equation}
G(t_f)  =   e^{i \epsilon_0 t_f} \mbox{det}'\mid \sigma \mid
\label{eq:G=det_sigma} 
\end{equation}
while from (\ref{eq:motion}, \ref{eq:R(t)} and \ref{eq:R(lambda)})
\begin{equation}
R(t)  =  S(t) \quad  \mbox{and} \quad 
R(\lambda,t) = \exp{\lambda \log S(t)} . 
\label{eq:R(lambda,t)}
\end{equation}
The RH problem (\ref{eq:jump_condition} and 
\ref{eq:Y->1}) reduces to
\begin{equation}
Y_-(t)Y_+^{-1}(t) = S^\lambda(t)  \qquad Y \rightarrow 1 
\; \mbox{when} \; |t|\rightarrow \infty .
\label{eq:RH_for_G} 
\end{equation}
When the matrix $S$ is constant between $0$ and $t_f$,
we will denote its value by $S^e$.
In the single-channel case, $S^e =  e^{2i\delta}$
and the RH problem is solved by
\cite{Muskhelishvili}
\begin{equation}
Y(z) = \exp{\left[\frac{1}{2\pi i} \ln{\left(\frac{z}{z-t_f} \right)}
\lambda \log{S^e} \right] } .
\label{eq:solution_commuting_S}
\end{equation}
(This solution was used implicitly in the original solution
to the single channel problem of \cite{ND69}.)
In fact the result (\ref{eq:solution_commuting_S}) solves
the RH problem even where the problem is not separable provided
that the matrices $S(t)$ evaluated at different times 
$t$ with $0<t<t_f$ commute. (This can be checked by direct substitution
into (\ref{eq:RH_for_G}).)
We insert $Y(z)$ and $R(\lambda,t)$
into (\ref{eq:rhtime-trace}). The integral over $t$
runs between 0 and $t_f$ where $\log{R(\lambda)}$ is non-zero.  
Inserting into (\ref{eq:log_chi}) and including
the factor of $e^{i\epsilon_0t_f}$  yields
\begin{equation}
\log{\chi_R} = 
i\epsilon_0' t - 
\log{i\xi_0 t_f} \left(\frac{\delta}{\pi} \right)^2 , 
\label{eq:log_chi_fes}
\end{equation}
where $\epsilon_0'=\epsilon_0 + \sum_{\epsilon<0} \delta(\epsilon)/\pi 
\nu(\epsilon)$, with $\nu(\epsilon)$ the density of states, is the
shifted energy of the core-hole in the presence of the Fermi gas. 
(The
form for the difference between $\epsilon_0$ and $\epsilon_0'$ is usually
attributed to Fumi \cite{Fumi55,Friedel52}.)
Close to the branch points of $Y$ at 0 and $t_f$, we 
cut the integrals off
at $i\xi_0^{-1}$ and $t_f+i\xi_0^{-1}$   
where $\xi_0$ is an energy of order the
band width.
Eq (\ref{eq:log_chi_fes}) gives  the well-known result for the long-time
asymptotic behavior of $G$ \cite{Mahan67}:
\begin{equation}
G(t_f) \sim (i\xi_0t)^{-\alpha} e^{i\epsilon_0't_f}, \qquad \alpha = (\delta/\pi)^2.
\label{eq:G(tf)}
\end{equation}

To compute the function $F(t_f)$ in (\ref{eq:open-line}) is
slightly more involved, although the underlying RH problem
is the same.
As already mentioned, the role of the core hole is to switch on the potential 
at $t=0$ and switch it off again at $t_f$, so that $F(t_f)$ can be written
(writing out the channel indices explicitly)
\begin{eqnarray}
F(t_f) & = &   \sum_{i\epsilon,i'\epsilon'}
W_{i\epsilon}^* 
\langle 0|  \hr(i\epsilon,i'\epsilon') | 0 \rangle
W_{i'\epsilon'} 
\label{eq:F_from_r} \\
\hr(i\epsilon,i'\epsilon') & = &
\hU_0^\dagger(t_f) \ha_{i\epsilon}  \hU (t_f)  \ha_{i'\epsilon'}^\dagger.
\label{eq:r(a,a)}
\end{eqnarray}
In the basis of the scattering states $\ha_{j'\alpha'}^+ |\rangle$
the matrix elements of this operator are easily shown to be given
in terms of $\sigma$ in (\ref{eq:motion}) by
\begin{equation}
r(i\epsilon,i'\epsilon')_{j\alpha  j'\alpha'}  =
e^{i(\epsilon_0-\epsilon) t_f} \left(
\sigma_{j\alpha j'\alpha'}\sigma_{i\epsilon i'\epsilon'}
-
\sigma_{j\alpha i'\epsilon'}\sigma_{i\epsilon j'\alpha'}
\right).
\label{eq:new_one_body}
\end{equation}
Using (\ref{eq:det}) and (\ref{eq:F_from_r}) we find
\begin{equation}
F(t_f) = e^{i\epsilon_0t_f}  \mbox{det}'\left| \, C \sigma 
- | h \rangle \langle g | \, \right| .
\label{eq:det'_open_line}
\end{equation}
Here $C=C(t_f)$ is the number:
\begin{equation}
C = \sum_{i\epsilon,i'\epsilon'} e^{-i\epsilon t_f}
W^*_{i\epsilon}  \sigma_{i\epsilon i'\epsilon'} W_{i'\epsilon'}
\label{eq:constant_C}
\end{equation}
while
\begin{eqnarray}
|h \rangle &  = & \sum_{j \alpha} 
\left(  
\sum_{i'\epsilon'} \sigma_{j\alpha i' \epsilon'}
W_{i'\epsilon'} \right) a_{j \alpha}^{\dagger} \mid \rangle 
\nonumber \\ 
\langle g|  &  = & \sum_{j' \alpha'} 
\left(\sum_{i\epsilon}e^{-i\epsilon t_f} W_{i\epsilon}^* 
\sigma_{i \epsilon j'\alpha'} \right) \langle | a_{j' \alpha'}  .
\label{eq:h_+_g}
\end{eqnarray}

The expression  (\ref{eq:det'_open_line}) is now in the form
(\ref{eq:det}). We could attempt to solve the corresponding
RH problem (\ref{eq:jump_condition}) and  (\ref{eq:Y->1}) as before,
although the relation between the corresponding operator $\hR(t)$
and $S$ is no longer simple.  
However, it is easier to simplify (\ref{eq:det_open_line})
by factoring out $G(t_f)=e^{i\epsilon_0t_f}\mbox{det}'|\sigma|$:
\begin{equation} 
F(t_f) =   C 
G(t_f)
\; \mbox{det} \left| \,  1 -  C^{-1} O | h \rangle \langle g | \, \right|
\label{eq:det_open_line}
\end{equation}
with
\begin{equation}
O  =  (1-f + f\sigma)^{-1} f .
\label{eq:operator_O}
\end{equation}
We have used (\ref{eq:full_det})
to put (\ref{eq:det_open_line}) in the form of  
the determinant over all states in the basis. Using the identity
$\mbox{det} |1 -  C^{-1}O  | h \rangle \langle g || = 
1 -  C^{-1}\langle g | O  | h \rangle $ we obtain
\begin{equation}
F(t_f) = G(t_f) \left( C -  \langle g | O  | h \rangle \right) .
\label{eq:F(t_f)_formula}
\end{equation}
As $C \rightarrow 0$ for large $t_f$ (with 
a functional form which depends on assumptions
about the density of states at the band edge),  
the response is determined by the second term.

The function $Y(z)$ with $\lambda=1$ in (\ref{eq:RH_for_G}
and \ref{eq:inverse_(1-f+fR)})
can be used  to compute $F(t_f)$.
In the time-representation
\begin{equation}
(1-f + f\sigma)^{-1} f =
Y_+ f Y_-^{-1} .
\label{eq:YfY}
\end{equation}
Writing $F(t_f)=L(t_f)G(t_f)$ ($L$ is
usually referred to as the open-line contribution), we find 
\begin{eqnarray}
L & = & - \sum_{l l'} \int d\epsilon d\epsilon' 
\int dt_1 dt_2
W_{l \epsilon}^*
e^{i\epsilon(t_1-t_f)}  \times  \nonumber \\
 & & \hspace{1em} 
\sqrt{\nu_l} \left[S Y_+ f Y_-^{-1} 
S \right]_{lt_1,l't_2}  \sqrt{\nu_{l'}} e^{-i\epsilon't_2} W_{l' \epsilon'}
\label{eq:L=WSYfYW}
\end{eqnarray}
Taking $W_i$ to independent of $\epsilon$ (we are assuming that the
long $t_f$ behavior is determined by states with energies
within $\sim 1/t_f$ of the Fermi surface), 
this simplifies to give
\begin{equation}
L \simeq W_{i}\sqrt{\nu_i} 
\left[ Y_-(t_f^-) \frac{1}{it_f} Y_+^{-1}(0^-) \right]_{ii'}
\sqrt{\nu_{i'}} W_{i'} .
\label{eq:L=YfY}
\end{equation}
The functions $Y_-$ and $Y_+$ are evaluated at $t=0^-$ 
and $t=t_f^-$. This prescription is equivalent to 
the imaginary time cutoff
used to derive  (\ref{eq:log_chi_fes}) and used 
in \cite{ND69,Matveev-Larkin}. 
Strictly, the discontinuities in $S$ at $t=0$ and 
$t=t_f$ should be thought of as the limit of a fast switching process,
in which $S$ starts to change at $t=0$ and reaches its new value
$S^e$ after a short time. 
Similarly at $t=t_f$, $S$ starts to change back from
$S^e$ to its unperturbed value.
(The corrections associated with a more 
realistic model of a non-instantaneous switching
process were considered for a related problem in \cite{AM01}.) 
In the single channel case we can insert the explicit
form for $Y$ given by (\ref{eq:solution_commuting_S}),
and recover the standard results
\begin{equation}
L \sim \frac{1}{it_f} \frac{1}{ (i\xi_0 t_f)^{-2\delta/\pi} },
\; F \sim \frac{1}{it_f} 
\frac{1}{ (i\xi_0 t_f)^{(\delta/\pi)^2-2\delta/\pi}}.
\label{eq:L_single-channel}
\end{equation}
 
When the potential $M(t)$ in (\ref{eq:H(t)}) is strong enough
for a bound state of the Fermi gas electrons to form below
the bottom of the band, the results for $G(t_f)$
and $F(t_f)$ are no longer correct. Provided the perturbing
potential is constant between $0$ and $t_f$, the effect
of the bound state can be taken into account explicitly
as explained in the appendix. The result
for $G(t_f)$ given by (\ref{eq:G+bound}), and 
for $F(t_f)$ given by (\ref{eq:open_line+bound}),
have two main contributions. After taking the Fourier transform 
to obtain the absorption spectrum, the first corresponds to having 
the bound state occupied and leads to the absolute threshold for 
absorption. The
second term relates to scattering processes in which
the bound state is always empty and leads to a subsidiary
threshold at $E_b$ above the first in the absorption spectrum.

The results (\ref{eq:G(tf)}), (\ref{eq:L_single-channel}),
(\ref{eq:open_line+bound_single_channel})
are of course very well-known \cite{ND69,Schotte+Schotte69}. However, 
none of the key formulas
(\ref{eq:solution_commuting_S}, \ref{eq:L=YfY} and
\ref{eq:open_line+bound}) require
that the scattering matrix $S$ should be diagonal in the
channel indices. Provided that $S(t)$ commutes with itself at
different times, the results are valid for arbitrary channel number.
We can therefore use the function $Y(z)$ given by
(\ref{eq:solution_commuting_S}) to compute the corresponding
results for the case of a non-separable potential just 
as easily as in the separable case.
In the absence of bound states, one obtains with
$\epsilon'_0 = \epsilon_0 + \sum_{\epsilon < 0} 
\sum_\zeta \delta_\zeta(\epsilon)/\pi \nu_\zeta(\epsilon) 
$:
\begin{eqnarray}
G(t_f) & = & \exp{(i\epsilon'_0 t_f)} \; (i\xi_0 t_f)^{-\beta}
  \nonumber
\\
L(t_f) & = & \sum_\zeta |\tau_\zeta|^2 
\frac{1}{it_f} 
\exp{\left(\frac{2}{\pi} \delta_\zeta \ln{i\xi_0 t_f} \right) } \nonumber \\
\beta & = & \sum_\zeta\left(\frac{\delta_\zeta}{\pi} \right)^2 ,
\label{eq:G+L_2ch}
\end{eqnarray}
which are the results obtained perturbatively  in \cite{Matveev-Larkin}. 
Here the eigen values of the matrix $S^e$ (see after
\ref{eq:RH_for_G}) are written as $e^{2i\delta_\zeta}$.
$S^e$ has eigen vectors $f(\zeta)_i$ 
and $\tau_\zeta = \sum_i \sqrt{\nu_i} W_i^* f(\zeta)_i$.

The perturbing potential, characterised by scattering
matrix $S^e$, can be strong enough to lead
to a bound state with wavefunction given
by (\ref{eq:bound_state}). 
In the presence of a bound state(s) we take the eigenvalues of 
$S^e$ to be $e^{i2\tilde{\delta}_\zeta}$ with $\tilde{\delta}_\zeta$
defined as the phase shift modulo $\pi$ in
channel $\zeta$ on the interval
$[-\pi/2, \pi/2]$ (see discussion after \ref{eq:bAb_single_channel}).  
We then obtain the generalizations to the non-separable case of the results 
of \cite{CN71} for $G(t_f)$ and $F(t_f)$. We find 
\begin{equation}
G(t_f)  =  \tilde{G}(t_f)\left(1 +  A_B \right),
\label{eq:G+bound2}
\end{equation}
where $\tilde{G}(t_f)$ is the contribution of the scattering
states given by the expression for $G$ in (\ref{eq:G+L_2ch})
with phase shifts given by $\tilde{\delta}_\zeta$, while
\begin{equation}
A_B \sim e^{-iE_B t_f}\sum_\zeta |\eta_\zeta|^2 e^{-2i\tilde{\delta}_\zeta}
\exp{\left(-\frac{2}{\pi} \tilde{\delta}_\zeta \ln{i\xi_0 t_f} \right) }.
\label{eq:bAb_multi_channel}
\end{equation}
Here $\eta_\zeta = \sum_i \sqrt{\nu_i} u_i^* f(\zeta)_i$ 
and the $u_i$ are the bound state wavefunction coefficients 
given in (\ref{eq:bound_state}).
In the presence of the bound state, the function 
$F(t_f) \sim F_0(t_f) + F_b(t_f)$ with
\begin{eqnarray}
F_0(t_f) & \sim & \tilde{G}(t_f) \tilde{L}(t_f) \nonumber \\
F_b(t_f) & \sim & e^{-iE_B t_f} \tilde{G}(t_f) |\bu \cdot \bW|^2,
\label{eq:F_bound}
\end{eqnarray}
where $\tilde{L}(t_f)$ is the scattering state contribution
to $L(t_f)$ given by the expression in (\ref{eq:G+L_2ch}), using
the phase shifts $\tilde{\delta}_\zeta$.

The formulas (\ref{eq:F_bound}) and (\ref{eq:G+L_2ch}) are the
natural generalizations of the single channel result and have
exactly the same interpretation as was given originally
in \cite{Schotte+Schotte69,Schotte_bosons69}. We repeat this briefly
here as the results for the non-equilibrium case (given in the
next section) can also
be understood heuristically on a similar basis but with
the phase shifts becoming complex. The exponents $(\delta_\zeta /\pi)^2$
and $(\delta_\zeta/\pi \pm 1)^2$ are, according to the Friedel sum rule,
the square of the net charge that needs to move in to or away from the origin
in order to screen the core hole potential. For $G(t_f)$ this is
$\delta_\zeta/\pi$, while for $F(t_f)$ it is $(\delta_\zeta - \pi)/\pi$ if
the photoelectron inserted at the origin is in the $\zeta$ channel
and $\delta_\zeta /\pi$ otherwise. If there is  an occupied bound state
after absorption of the photon,
the respective values become $(\tilde{\delta}_\zeta + \pi)/\pi$ and 
$\tilde{\delta}_\zeta/\pi$, as now the 
Fermi gas has to provide the additional electron which ends
up in the bound state. The form $t^{-n^2}$ is just the
decay with time of the overlap of the wavefunction of
the Fermi gas at $t=t_f$ and the one describing the 
system created at $t=0$ in which (with respect to the 
ground state in the presence of the core hole) 
there is an excess charge $n=-\delta/\pi$ at the origin.
 That it vanishes as $t\rightarrow \infty$, is the 
orthogonality catastrophe described by Anderson \cite{Anderson67}.

\subsection{Non-equilibrium Effects}

The  experimental and technological interest
in the out-of-equilibrium response of coupled Fermi systems
has grown as electronic devices have shrunk. 
Examples include structured quantum dots, like the single
electron transistor or the single photon detector \cite{Komiyama00}, 
and quantum point contacts.
The non-equilibrium Fermi Edge Singularity (nFES) will characterize 
the  energy absorbed by the coupled  Fermi gases 
in a rapid switching process in such devices.
The nFES should help explain, for example,
measurements of random telegraph signals (RTS).
In these experiments, a two-level system (TLS) couples to the
source-drain current flowing in the channel of a MOSFET (the TLS resides
in the insulating oxide layer \cite{Cobden-Muzykantskii95}). 
The RTS relates to the `random' switching
of the TLS between its ground and excited states.  The ratio between the
times the TLS spends in the excited and ground states is measured
experimentally.  In equilibrium this ratio is
fixed by detailed balance, and the deviations from this
have been attributed to non-equilibrium effects \cite{CobdenRTS-unpubl}.

As one of the simplest non-trivial many-body effects, the FES
is also a natural point to start, when looking for a description
of non-equilibrium effects in many-electron systems. 
Perhaps surprisingly, given its conceptual simplicity, 
the nFES has not attracted as much attention
as more difficult non-equilibrium problems like the Kondo
effect, to which it is known to be related. (The Kondo effect
can be thought of as a sequence of FES's
associated with each flipping of the localized moment
\cite{Anderson69}.)

\begin{figure}
{\centering	
\input{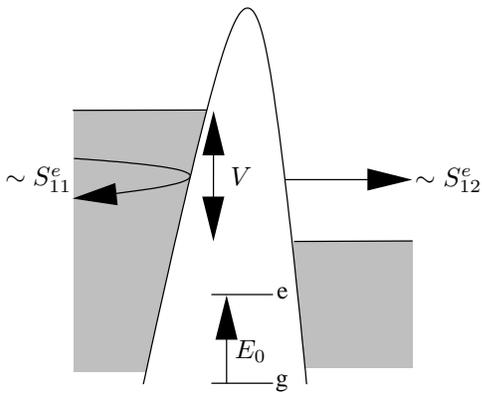}
 }
\caption{Energy levels in an 
idealized device to demonstrate the out-of-equilibrium
FES. The scattering potential for electrons is characterized via the
$2\times2$  matrix,  $S(\epsilon)$,  connecting scattering states in the
two wires for particles with energy $\epsilon$.
$S=S^g$ or $S^e$ depending on whether
the defect is in its ground ($g$) or excited ($e$) state
(with excitation energy $E_0$). 
$S^g$ is the identity matrix and $S^e$
is an arbitrary unitary matrix. $S^e_{11}$ and $S^e_{12}$ correspond
to the reflection and transmission amplitudes respectively.
We will refer to the device operating as illustrated here, 
with a negative potential $-V$ ($V>0$)  
applied to the left electrode, as the 
forward-biased case.
\label{fig:device}}
\end{figure}

In \cite{MdAB03} we reported results for
$G(t_f)$ for a two-channel problem, which modelled a
system with two electrodes separated by a barrier. The
transmission of the barrier depends  on the state
of a two-level system inside the barrier, see Figure 
\ref{fig:device}, with the transition between the two-levels
assumed to be dipolar. The real part of the Fourier transform
of the function $G(t_f)$ gives the 
absorption spectrum for the device.
The non-equilibrium effects predicted in \cite{MdAB03} should be visible
in the voltage dependence of the absorption line-shape of
devices like the single-photon detector of  \cite{Komiyama00}
and \cite{Shields_etal00}.
In \cite{Komiyama00}  a quantum dot in the 
quantum Hall regime is coupled via tunneling barriers to
two electrodes  on either side of the dot. 
For magnetic fields in the range 3.4-4.2T, the conductance
through the dot can change from  zero to around
0.3 $e^2/h$ when a photon is absorbed via cyclotron
resonance in the dot. From the perspective of the
two electrodes, the  dot behaves as a tunneling
barrier which allows tunneling only in its excited state.
The absorption of the photon and
the subsequent separation of the hole (which 
moves into the $\nu=1$ ring on the outer part of the dot)
and the particle (which  `falls' into center of 
the dot at $\nu=2$) is rapid, while the response
of the conduction electrons in the two electrodes 
is slow and will show effects characteristic of the
FES. In the device of \cite{Shields_etal00}, a electron
trapped in a dot underneath
an electron channel gives rise to a potential which closes
off a conducting channel. When a photon is absorbed, the photo-exicted
hole can recombine with the electron in the trap, 
the potential of the electron disappears and the channel opens. Again 
the conduction electrons on the two sides of the
channel, `see' the sudden reduction of a tunneling
barrier on absorption of a photon.

The main result reported in \cite{MdAB03} was that
the ND formula describing the form of
the FES and threshold shift (Fumi's theorem \cite{Fumi55,Friedel52})
generalized in a simple way to the non-equilibrium case.
For time scales $t\ll t_f$, the
phase shifts which appear in $G(t_f)$ are real
and are given by the $\log$ of the eigen-values of the scattering 
matrix $S^e$ \cite{YY82,Matveev-Larkin}. This simply
reflects the fact that on these short time scales
the response of the Fermi gas involves excitations with
energies $\epsilon \gg V$ which do not sense the non-equilibrium
distribution function.
On time scales $t \gg t_f$, 
the equilibrium phase shifts in the two channels are 
replaced by
`complex' phase shifts given by $\log{S^e_{11}}$
and $\log{(1/S^e_{22})^*}$. The real part of these phase
shifts describes the scattering within each electrode,
while the imaginary part describes the effect of
scattering processes in which particles cross
the barrier.
One effect of the non-equilibrium operation of
the device is to make the scattering between
the different electrodes effectively incoherent. Here, we find that 
this interpretation extends also
for the function $F(t_f)$.

We show the key steps in the derivation of 
$G(t_f)$, emphasizing the relationship with the
equilibrium results, and report the results for 
$F(t_f)$ including the role of possible bound states.
Since the initial state involves a filled Fermi
sea in both channels (left and right electrodes),
the RH formulation of this non-equilibrium problem
is the same as for the equilibrium case. 
The  bias across the tunnel junction means
only that the chemical
potentials are different in the two electrodes. One
way of handling this difference is to introduce
a gauge transformation acting only on the basis states
in the left electrode: 
\begin{eqnarray}
\mathbf{a}(\epsilon) & \rightarrow & \mathbf{a}(\epsilon,t) =  
e^{+iL\int_0^tV(\tau)d\tau}
\mathbf{a}(\epsilon) 
\label{eq:gauge_transform} \\ 
S(t) & \rightarrow & S(t)=e^{+iL\int_0^tV(\tau)d\tau} S(t) 
e^{-iL\int_0^tV(\tau)d\tau} 
\label{eq:S(t,lambda)}
\end{eqnarray}
where $L$ is the diagonal matrix with $L_{11}=1$ and  $L_{22}=0$.
The effect of this transformation on states
in the left electrode is to set $\epsilon \rightarrow \epsilon - V(t)$, 
so that  the chemical potential  in the left electrode 
becomes equal to that in the right electrode (taken to be zero as before). 
For the constant bias case, the transformation  gives
$\ba(t) \rightarrow \ba(t)=e^{+iLVt} \ba $.

The functions $\log \chi_R^{(2)}$ and $L(t_f)$ for the
nFES case are still given
by (\ref{eq:rhtime-trace}) and (\ref{eq:L=WSYfYW}).
However,
the RH  problem satisfied by the function $Y(z)$ is different:
In  (\ref{eq:RH_for_G})  $S^\lambda(t)$  picks
up an additional time-dependence from the gauge
transformation (\ref{eq:S(t,lambda)}), which leads to
two important differences to the equilibrium case. Firstly,
the function $e^{iVt}$ introduces a new characteristic 
energy scale, $V$. If the function $S(t)$  has Fourier components 
with freqencies $\omega \gg V$, the response will be dominated 
by states with energies $|\epsilon| \gg V$ and will be
insensitive to the non-equilibrium nature of the 
distribution which only becomes 
apparent on the energy scale $V$. If $S(t)$ only varies
at frequencies $\omega \ll  V$ the response will come from
states with energies $\epsilon \ll V$ and will normally
be significantly different from what happens in equilibrium.

The second main difference following from the 
additional time-dependence of $S(t)$ relates to the case when
between $t=0$ and $t=t_f$
the scattering matrix (before the gauge transformation)
is constant and equal to $S^e$. In this case it is now no longer
possible  to solve the RH problem
with a function of the form (\ref{eq:solution_commuting_S}). 
Although this form satisfies formally the jump condition, 
$Y_-(t)Y_+^{-1}(t) = S^\lambda(t)$, the corresponding function $Y(z)$
is not well-defined for large $z$ if $S^e$ is not
diagonal. The off-diagonal elements of $S^e$ contain factors
$e^{\pm iVt}$ so that there is an 
essential singularity at $z \rightarrow \infty$ in
$Y(z)$ defined by (\ref{eq:solution_commuting_S}), and
it no longer satisfies the condition $Y \rightarrow 1$.
This problem is clearly apparent in the RH formulation we have
presented. It was much less clear 
in previous attempts  to extend the ND method to the
non-equilibrium case and may explain why these failed \cite{CR00}.
It is also interesting to note that for $t_f \ll 1/V$, 
we can expand the function $e^{iVt}$ up to linear
order in $Vt_f$. Then $S^e = S^e(V=0) + CVt$, 
there is no singularity at infinity for $Y$,
and the form (\ref{eq:solution_commuting_S}) 
still works.

In general there is no exact solution to the non-commuting
RH problem \cite{DeiftZhou}.
However, in the case relevant to the device
shown in Fig \ref{fig:device},
$S(t)= e^{iLVt} S^e e^{-iLVt}$ for 
$0<t<t_f$, with $S^e$ constant,
we can find an asymptotically correct solution 
for the limit $t_f \gg 1/V$ relevant
to the nFES \cite{MA03,MdAB03}. 
We will only consider the case where there is
one channel in each electrode.
As in the equilibrium case (cf. \ref{eq:R(lambda,t)})
\begin{equation}
R = \exp{ (\lambda \log{S(t)} ) } .
\label{eq:RnFES}
\end{equation}
In this case the 
solution for $Y(t)$, valid for $t \gg V^{-1}$, is given
for $t<0$ or $t>t_f$ by
\begin{equation}
Y(t,\lambda)  =   
     \psi(t,\lambda), 
\label{eq:Y_region1+4}
\end{equation}
while above and below the cut, $[0,t_f]$,
\begin{eqnarray}
Y_+(t,\lambda) & = &       
      \begin{pmatrix}
        1 & -\gamma(t,\lambda)   \\ 
        0 & 1
      \end{pmatrix} \psi_+(t,\lambda) 
\nonumber \\
Y_-(t,\lambda) & = &      
      \begin{pmatrix}
        1 & 0  \\ 
        +\eta(t,\lambda) & 1
      \end{pmatrix} \psi_-(t,\lambda) . 
\label{eq:Y_large_t} 
\end{eqnarray}
Here $\gamma(t,\lambda)=R_{12}/
R_{11}$ and $\eta(t,\lambda)=R_{21}/R_{11}$. The functions
$\psi_\pm (t,\lambda)=\psi(t\pm i0,\lambda)$, where
$\psi(z,\lambda)$ is given by
\begin{equation}
  \psi  =   
 \exp{\left[ 
\left(  x_1 \tau_0 + x_2 \tau_3    \right) 
\log \frac{z}{z-t_f} \right]} ,
\label{eq:psi}
\end{equation}
with
\begin{equation}
x_1(\lambda)  =  \frac{ \log{ { 
R_{11} / R^*_{22}} } }{4\pi i} 
\quad \mbox{and} 
\quad x_2(\lambda) = \frac{\log{{
(R_{11}R^*_{22})}}}{4\pi i} .
\label{eq:x1x2}
  \end{equation}
Here  $\tau_3 $ is the third Pauli spin matrix and 
$\tau_0$ is the identity matrix.
The derivation of (\ref{eq:Y_large_t}) follows that given
in \cite{MA03}. The idea,
which was explained in detail 
in the context of inverse scattering problems in \cite{DeiftZhou},
is to solve for a function $W(z)$ which satisfies the same
jump condition as $Y(z)$ but in a complex plane with additional
cuts. For this problem, the additional cuts are parallel to the imaginary axis
and run from the branch points at $z=0$ and $z=t_f$ to infinity. 
The discontinuities in $W(z)$ across the vertical cuts 
scale as $e^{-|Vz|}$. If $Y$ is approximated by $W$, the errors
in $\log{\chi_R}$ defined in (\ref{eq:log_chi}) are only $O(1/Vt_f)$,
and can, in principle, be computed order by order in powers of
$(Vt_f)^{-1}$. 

The form for $\log \chi_R$ for $t_f\gg V^{-1}$ is found by
inserting (\ref{eq:Y_large_t}) into (\ref{eq:rhtime-trace})
and (\ref{eq:log_chi}) and computing the integrals over
$t$ and $\lambda$ as in the equilibrium case \cite{MdAB03}:
\begin{equation} 
\log{\chi(t_f,V)}   =  -i(E_0-\Delta(V)) t_f - \beta' \log{(iVt_f)} + D,
\label{eq:Vneq0}  
\end{equation}
where $\Delta(V)$ is given by the non-equilibrium generalization
of Fumi's theorem \cite{Fumi55,Friedel52}
\begin{equation}
\Delta(V)  =  \int_{-\infty}^0  \frac{\mbox{tr}\log{(S^e(E))}}{2\pi i} dE + 
\int_0^V \frac{\log{(S^e_{11}(E)) }}{2\pi i} dE . 
\label{eq:Delta(V)} 
\end{equation}
The constant $\beta'$ is given by (cf.\/ \ref{eq:G+L_2ch}):
\begin{equation}
\beta' = \left( \frac{\log{(S^e_{11})} }{2\pi i} \right)^2
+ \left( \frac{\log{(1/S^e_{22})^*} }{2\pi i} \right)^2 .
\label{eq:beta'}
\end{equation}
The constant term $D$ can be estimated by requiring that the
form for $\log \chi$ (\ref{eq:Delta(V)}) matches the 
equilibrium one at $t_f=V^{-1}$ 
(\ref{eq:log_chi_fes}) valid for
$t_f \ll V^{-1}$. This constant
gives the contribution from excitations
with frequencies between $V$ and $\xi_0$. This gives:
\begin{equation}
D = \beta \log{\xi_0/V}.
\label{eq:D}
\end{equation}

\begin{figure}
{\centering \input{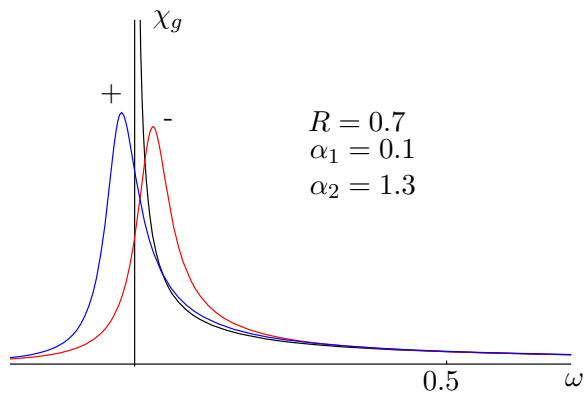} }
  \caption{Spectral function
$\mbox{Re}\chi_G(\omega)$ computed from (\ref{eq:chiE}) with
$\omega$ in units of the bias voltage $V$.
The spectra depend on $S^e_{11}=\sqrt{R}e^{i2\alpha_1}$ and 
$S^e_{22}=\sqrt{R}e^{i2\alpha_2}$, where $R$ is the reflection
probability. The curve marked
$+$ ($-$) refers to the case in which electrode 1 (2) is at
the higher chemical potential. Also shown is the corresponding equilibrium
result calculated from (\ref{eq:G+L_2ch}) using $\xi_0 = V$
(\cite{note_on_freqs}).  In addition to the overall
smoothing of the singularities, expected in a non-equilibrium system, 
there are two significant non-equilibrium
features. Firstly, the maximum in the spectral weight
is shifted away from its equilibrium value by an amount proportional
to the applied voltage. The shift,  Re$(\Delta(V) - \Delta(0))$,
which is given in the forward-biased case
in (\ref{eq:Delta2(V)}), depends on the polarity of the voltage.  Secondly
the form of the function changes on reversing the polarity of the 
device. 
}
  \label{fig:Gresponse}
\end{figure}

The result for $G(t_f)$ can be seen as an adaptation of the
equilibrium result.
The real phase shifts  (given by
$-i$ times the logarithms of the eigenvalues of the scattering matrix
$S^e$), which appear in the formulas (\ref{eq:G+L_2ch}), are
replaced by complex phase shifts. In the forward bias case described
by (\ref{eq:G+L_2ch}), these are $-i\log{S^e_{11}}$ and 
$-i\log{(1/S^e_{22})^*}$. The effect of the complex phase shifts
is to smooth the singularity seen in equilibrium (this could be expected
on quite general grounds) and to introduce a polarity dependence.
This polarity dependence affects both the shape and the position of the
spectrum and is evident in Figure \ref{fig:Gresponse}
where we show $\chi_G(\omega)$ for a particular choice of $S^e$. 
The dependence of the spectrum, $\chi_G(\omega)$, 
on the polarity of the device, when operating out of equilibrium, is governed 
by the difference $\alpha_{12} \equiv \alpha_1 - \alpha_2$ (with $\alpha_{1,2}$ as 
defined in the figure caption). The difference in the overall position of the spectrum
on changing the polarity is given by the difference in the second term on
the right hand side of (\ref{eq:Delta(V)}) and is proportional to $\alpha_{12}$.
This origin of this shift of the spectrum is the change in the nature of
the  scattering across the barrier from fully coherent in the equilibrium case to incoherent
for times $t_f \gg V^{-1}$ in the non-equilibrium case. 
The shape of the spectrum reflects the decay of charge from its
initial distribution (the equilibrium distribution for $S = 1$)
to the steady-state distribution for $S=S^e$ \cite{Schotte+Schotte69,Schotte_bosons69}.
In the non-equilibrium case, this decay can occur differently depending on the polarity.
If more charge is needed in the left-hand electrode to
screen the potential characterized by $S^e$, than in the right-hand one ($\alpha_{12} > 0$), 
this charge can come from states within $V$ of the Fermi energy of the 
right-hand electrode when the device is reverse-biased but not when it is forward-biased.

\begin{figure}[tbp]
{\centering \input{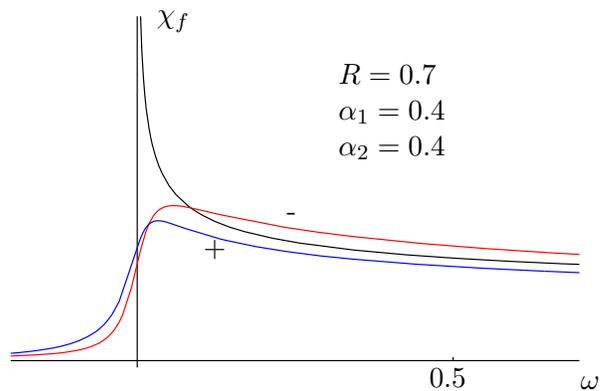} }
  \caption{Spectral function $\mbox{Re}\chi_F(\omega)$
computed from (\ref{eq:chiE} with
$\omega$ in units of the bias voltage $V$ for the case 
$(W_1,W_2) \sim (0,1)$ in (\ref{eq:L=YfY}).
The curve marked $+$ ($-$) refers to the case in which electrode 1 (2) is at
the higher chemical potential. Also shown is the corresponding equilibrium result
\cite{note_on_freqs}. For these relatively small phase shifts the 
singularity seen in equilibrium  disappears completely, although there
is still a polarity dependence of the spectrum even though the scattering
matrix is symmetric.
}
  \label{fig:Fresponse}
\end{figure}

For the model device shown in Fig \ref{fig:device} the 
absorption spectrum is given by the Fourier transform
(see Appendix \ref{sec:spectrum})
of $G(t_f)$ rather than $F(t_f)$,
as the transition in the barrier is presumed
to be dipolar. However, the corresponding function
$F(t_f)$ is also important. In   \cite{YA70,Anderson_Yuval_Hamann70}
Yuval and Anderson showed that the Kondo problem could be treated 
as an infinite sequence of spin flips or switching events,
with the response of the conduction  electrons
to each switching event characterized by  $F(t_f)$. Given
the long-standing interest in non-equilibrium effects
in the Kondo effect \cite{dAW82,Paaske03}, the 
correct non-equilibrium form for $F(t_f)$ would be the
starting point for the study of the non-equilibrium
Kondo effect using a generalization of the Yuval-Anderson
mapping. 

We must first write the function $F$ and the open
line function $L$ in terms of the
gauge-transformed basis:
\begin{equation}
L(t_f) \simeq W_{i}^* \sqrt{\nu_i} e^{-iLVt_f} 
\left[ Y_-(t_f) \frac{1}{it_f} Y_+^{-1}(0) \right]_{ii'} 
W_{i'} \sqrt{\nu_{i'}}.
\label{eq:L_n=YfY}
\end{equation}
We can now insert the solution for $Y_+$ (with $\lambda=1$)
from (\ref{eq:Y_large_t}) into  (\ref{eq:L=YfY}). The result
can be written:
\begin{equation}
L(t_f)  \simeq  \frac{1}{it_f}  \bW^*
      \begin{pmatrix}
        z_f (iVt_f)^{2x_+} & 0  \\ 
        \alpha z_f (iVt_f)^{2x_+}  & (iVt_f)^{2x_-}
      \end{pmatrix}  \bW,
\label{eq:L_n} 
\end{equation}
where $x_\pm = x_1(1) \pm x_2(1)$ with $x_i(\lambda)$ defined in
(\ref{eq:x1x2}), $\alpha = S^e_{21}/S^e_{11}$ and $z_f=e^{-iVt_f}$
\cite{note_on_zf}. 
The absence of a contribution proportional to 
$W_1^* W_2 z_f$ is to be expected. This would involve a contribution
to the open-line function 
from an electron initially placed in the right-hand electrode exciting
the Fermi gas in the left electrode. Since we are 
assuming that the tunneling
through the barrier by the electron 
is a slow process on the scale of $1/V$, this
does not lead to a singular contribution to $F$. (There is still a
contribution to $F$ proportional to
$W_1^* W_2 z_f$ from  the direct scattering term  $CG(t_f)$ 
in \ref{eq:F(t_f)_formula}.)

The effect of the open line contribution on 
$F(t_f)=G(t_f)L(t_f)$ is the natural generalization 
of the equilibrium result that one might expect given the
results for $G(t_f)$.  The 
corresponding spectral functions is shown in Figure \ref{fig:Fresponse}
for a particular choice of $S^e$. 
For simplicity we only look at the case where the
electron is added and removed from the same
($i=2$) electrode, {\it ie\/} $\bW \sim (0,1)$. 
The dependence of the spectrum on the polarity of the 
device, is present even in the case where $S^e$ is symmetric.
When a particle is added to an electrode,
the response of the system  will be depend on whether
the electrode is at the higher or lower  chemical potential.
The form of the spectrum can also differ substantially from
what happens in equilibrium.
For the case  $\alpha_1 = \alpha_2 = 0.4$ and $R=0.7$
shown in Figure \ref{fig:Fresponse}, there is no real peak left 
over from the  equilibrium result. This is because 
the phase shifts $\delta_\zeta$, corresponding to
the eigen modes of $S^e$, are small, and hence the
exponents in (\ref{eq:G+L_2ch}) for
the equilibrium 
function $\chi_F(\omega)\sim \omega^{-(\delta_\zeta-\pi)^2/\pi^2}$ 
are also small. The corresponding singularity is weak and easily
smoothed out by the finite lifetimes of states close to the Fermi
energy in the non-equilibrium case. This smoothing is enhanced because
one of the phase shifts, $\delta_\zeta$, is always larger than
$\alpha_1$ and $\alpha_2$. This larger exponent
gives the dominant singularity in equilibrium, but is
then effectively replaced by $\alpha_1$ out of equilibrium.
\section{Charge Transfer: (CSAC)}
The existence of Coherent States of Alternating Current (CSAC)
was predicted in \cite{IvaLL97}.
These consist of a sequence of pulses which propagate
through a contact. When the bias
across the contact is described by a class of
periodic (with period $\Omega$) rational
functions of the variable $z=e^{i\Omega t}$, then
the shot noise is minimised and the noise distribution does not 
depend on the separation of the pulses. 
This result is still 
not well understood, nor is it possible, using
the original derivation, to  establish how robust these states
are against deviations from 
zero temperature or from the ideal pulse shape.

Recent rapid experimental progress in the application of micro-wave
radiation at low temperatures suggests that the experimental test of the
existence of the CSAC is just about possible. Several experimental
groups are  pushing the technology in this
direction~\cite{AguK00,RosSS02,QinWT03}, and it should only be a matter
of time before experimental data becomes available.  However,
interpretation of these future experiments will not be easy using the
analytical method used in \cite{IvaLL97} as this depends crucially on the
particular shape of the pulses. There are no predictions about what
happens when the shape of the pulses deviates slightly from the required
one (something unavoidable in any real experiment), nor is the effect of
non-zero temperature known.

Here we show that the results of \cite{IvaLL97} for the 
CSAC are easy
to derive using the RH approach.  
When the pulses are periodic as in the case of the CSAC, the RH problem
simplifies. It requires solving for functions which 
are analytic in two {\bf disconnected} regions (inside and outside
the unit circle) with  the jump function specifying the discontinuity
across the boundary between
them. We show that the particular case of the CSAC 
corresponds to an RH problem which can be 
solved exactly using combinations of meromorphic functions in the plane---one
of which is analytic inside and one outside the unit circle.

The model device considered in  \cite{IvaLL97,MA03} consists
of a tunnel junction driven by a bias voltage $V(t)$ which
is periodic in time with period $T = 2\pi/\Omega$. It is 
equivalent to the device shown in Fig \ref{fig:device}.
We are interested in the change
in physical quantities over one cycle of the pump
in the limit $t_f \gg T$. In this limit,
effects induced by the switching on and off of the
periodic potential at $t=0$ and $t=t_f$ are 
irrelevant.
The scattering matrix $S$ is at some constant value
$S^{e}$ between $0$ and $t_f$. 
Applying the time-dependent gauge transformation 
(\ref{eq:S(t,lambda)}) leads to $S$ becoming a
periodic function of time, so that it
no longer commutes with itself at different times.

The distribution function for any single-particle observable
measured in this periodically pumped Fermi system will 
involve the solution of a non-commuting RH problem. In 
particular, 
the characteristic function or generating function for moments of the
distribution of the net transfer of charge from 
electrode 1 (left electrode)  to the  electrode 2 (right
electrode), $\chi(\lambda)$, is given by (\ref{eq:measurement})
with the operator $R$ given by
\begin{equation}
\hR(\lambda) = \hU^\dagger (t_f) e^{-i\lambda \hQ_1} \hU(t_f) e^{i\lambda \hQ_1}.
\label{eq:Rshot}
\end{equation}
Here 
\begin{equation}
\hQ_1 = \sum_\epsilon \ha_\epsilon^\dagger L \ha_\epsilon.
\label{eq:Q1}
\end{equation}
For states close to the Fermi surface ($E = \epsilon + \epsilon' \simeq 0$), 
the matrix $R$ in the time-representation can be written
\begin{equation}
R(t,\lambda) = S^{-1}(t) e^{-i\lambda L} S(t) e^{i \lambda L} ,
\label{eq:Rshot(t)}
\end{equation}
so that the characteristic function will be given by
(\ref{eq:full_det}, \ref{eq:log_chi}, \ref{eq:inverse_(1-f+fR)})
\begin{equation}
\log{\chi(\lambda)}  =  
\mbox{\rm Tr}\left(
\ln{( 1-  f +  f  R)} \right). 
\label{eq:log_chi_shot}
 \end{equation}

If the inverse of the solution $Y^+(t)$ to the RH problem 
(\ref{eq:jump_condition}, \ref{eq:Y->1})
with $R$ given by (\ref{eq:Rshot(t)}),
is analytic in the upper half-plane, we can write
the characteristic function as
\begin{equation}
\log{\chi(\lambda)} =   \int_0^\lambda d\lambda'
  \int dt\, \mbox{\rm tr}\left\{ 
Y_+fY_+^{-1}R^{-1} 
\frac{dR}{d\lambda'} \right\}
\label{eq:chi_shot_intermediate}
\end{equation}
Using (\ref{eq:Rshot(t)}) and (\ref{eq:dA/dt}), and computing
explicitly the derivative with respect to $\lambda$,
we obtain \cite{MA03,note_on_shot}
\begin{equation} 
\log{\chi(\lambda)} = 
 \int_0^\lambda \frac{d\lambda'}{2 \pi}
  \int dt\, \mbox{\rm tr}\left\{ 
\frac{d(Se^{i\lambda' L}Y_+)}{dt}(Se^{i\lambda' L}Y_+)^{-1}  \right\} .
\label{eq:shot_final}
\end{equation}
(If the eigenvalues of $Y_+$ 
have zeros in the upper half-plane, there are additional
contributions 
to the right hand side of the corresponding
relations to (\ref{eq:fYf_relations}) 
for $Y_+^{-1}$ from its poles. In this case,
$(1 - f + f R)^{-1}$ is no longer
given by (\ref{eq:inverse_(1-f+fR)}) but can be found using
methods described in Chapter 6 of \cite{Muskhelishvili}.)

In the case of the periodically driven
pump, the scattering matrix (after applying the gauge
transformation) is
periodic $ S(t)= S(t+T)$. If we change
variables to $z = e^{2\pi i t/T}$,  we need to find
a function $Y_+(z)$, which is analytic for $|z| < 1$ and $Y_-$ which
is analytic for  $|z| > 1$ and $ Y_- \to 1 \hbox{ when } |z| \to \infty $. 
On the unit circle $|z|=1$,
\begin{equation}
  \label{eq:RH-circle}
   Y_- Y_+^{-1}=S^{-1}(z) e^{-i\lambda L} S(z) e^{i \lambda L} .
\end{equation}
The characteristic function for charge transmitted during 
one cycle of the periodic pump in the limit
$t_f/T \gg 1$ is given by
\begin{eqnarray}
  \label{eq:chi-periodic}
  \log \chi &=&  
\int_0^\lambda \frac{d\lambda}{2\pi}
  \oint_{|z|=1} dz   \\
& & \mbox{tr} \left\{  \frac{d \left( S e^{i\lambda L} Y_+ \right) }{dz}  
 \left( S e^{i\lambda  L} Y_+ \right)^{-1} L \right\}  \nonumber
\end{eqnarray}

CSAC's were reported in  
\cite{IvaLL97} for the case when the phase factor
in (\ref{eq:S(t,lambda)})  can be written
as a  rational function, $l(z)$, of the variable $z=e^{2\pi i t/T}$:
\begin{eqnarray}
e^{+iL\int_0^tV(\tau)d\tau} & = & l(z) \nonumber \\
l(z) &  = & \prod_{i=1}^N \frac{z-a_i}{1- a_i^* z} ,
\label{eq:form_of_l(z)}
\end{eqnarray}
where either all $|a_i| > 0$ or $|a_i| < 0$. We can
choose $|a_i| > 0$ without loss of generality as 
$z \mapsto 1/z$ simply reverses the polarity of the
device.
In this case, we decompose $R(t,\lambda)$  (see \cite{MA03}) 
as follows
\begin{equation}
\label{eq:R-decomp}
R = \begin{pmatrix} 1& 0\\ \frac{\alpha}{l(z)} & 1 \end{pmatrix} 
     \begin{pmatrix} a& 0\\ 0 & \frac1a \end{pmatrix} 
     \begin{pmatrix} 1& \beta l(z) \\ 0 & 1 \end{pmatrix}, 
\end{equation}
where
\begin{eqnarray}  
a &=& |S_{12}|^2 e^{i\lambda} + |S_{11}|^2  \nonumber \\
\alpha &=& - \frac{S^e_{21} \left(S^e_{22} \right)^*
\left(1-e^{i\lambda}\right)}{a} \nonumber \\
\beta &=& - \frac{S^e_{12} \left(S^e_{11} \right)^*
\left(1-e^{i\lambda}\right)}{a}. 
\label{eq:a}
\end{eqnarray}
The solution to the RH problem
\begin{equation}
  \label{eq:RH-biased}
  Y_-Y_+^{-1}=
    \begin{pmatrix} 1& 0\\ \frac{\alpha}{l(z)} & 1 \end{pmatrix} 
     \begin{pmatrix} a& 0\\ 0 & \frac1a \end{pmatrix} 
     \begin{pmatrix} 1& \beta l(z)\\ 0 & 1 \end{pmatrix},
\end{equation}
is clearly
\begin{eqnarray*}
Y_-&=&\begin{pmatrix} 1&0\\ \frac{\alpha}{l(z)} & 1 \end{pmatrix} \\
Y_+&=&\begin{pmatrix} 1& -\beta l(z)\\ 0 & 1 \end{pmatrix} 
\begin{pmatrix} \frac1a & 0\\ 0 & a \end{pmatrix}.
\end{eqnarray*}   
Inserting this into (\ref{eq:chi-periodic}) gives
the result reported in \cite{IvaLL97}:
\begin{equation}
  \label{eq:contact-answer}
  \ln\chi(\lambda)= N \frac{t_f V}{2\pi}\ln a 
\end{equation}
with $a$ given by~(\ref{eq:a}).

The surprising feature of the result (\ref{eq:contact-answer}) 
is that it implies that the second moment of the shot noise 
$\ll  n^2\gg$ achieves the absolute minimum for given
charge transfer $<n>$ \cite{IvaLL97}, which is the value
obtained in the constant bias case $a_i \rightarrow \infty$ for
all $i$. The feature of the phase factor $l(z)$ which leads
to the RH problem being so easy to solve 
is that all its poles (zeros) are either inside or
outside the unit circle $|z| = 1$, which means that the decomposition
of $R$ in (\ref{eq:R-decomp}) automatically solves the 
RH problem. In the case of an arbitrary rational function for
$l(z)$ this is not the case as there can be points at
which $\mbox{det}|Y_+(z)|$ vanishes inside the unit circle.
The corresponding formulae to 
(\ref{eq:fYf_relations}) $Y_+^{-1}$ pick
up additional terms on the right hand side nad
$(1 - f + fR)^{-1}$ is not given by
(\ref{eq:inverse_(1-f+fR)}), although, in principle, it can still be found
given the solution to the RH problem $Y(z)$. 

\section{Conclusions and Outlook}
The RH approach is a general method for computing 
the response of a Fermi gas to a localized time-dependent perturbation.
There are two key steps to the method. First, 
provided the  condition 
(\ref{eq:slowness_condition}) is met, the method  works 
with the scattering matrix defined on the instantaneous
value of the potential rather than with the Hamiltonian.
This has the attractive feature of  working directly with
the physical quantities determining
the long time response of the system to a perturbation, namely
scattering amplitudes for particles close to the Fermi
surface.
The condition (\ref{eq:slowness_condition}) is essentially the
requirement that the perturbation varies more slowly than 
the delay time for a particle traversing the region in which 
the perturbation acts. The second key step is to relate
the response of the Fermi gas to the solution of a non-commuting RH problem
(\ref{eq:jump_condition}, \ref{eq:Y->1}).  
The RH problem corresponding to any given
experimental situation is usually easy to set up. Its solution
and the interpretation of the results is a more
delicate task which needs to be repeated for each new physical situation.
While there is no analytical solution of the general non-Abelian
RH problem,  there is a powerful technique for finding 
asymptotic solutions valid
for frequencies much smaller than those present
in the jump function \cite{DeiftZhou}. 

Here we have emphasized the generality of the approach and
applied it to two existing problems---the Fermi Edge
Singularity and the shot noise in a periodically pumped
tunnel junction. The calculations in the two cases are very
similar. In the case of the FES we have rederived all the known
results for the equilibrium case emphasizing, in particular,
how the method is no more complicated in the case of the 
non-separable potential than in the separable case. Our
derivation for the non-separable case is, we believe, the
first non-perturbative solution of this problem. 
For the non-equilibrium device shown in Fig \ref{fig:device}, we
have explained how the results for the core-hole Green's
function of  \cite{MdAB03} were obtained and given the
corresponding results for the open-line function $L(t_f)$ 
(\ref{eq:L_n}). For the case of the CSAC's, we have shown
that the particular form of the periodically varying 
bias with the phase factor $l(e^{2\pi i t/T})$ given
by (\ref{eq:form_of_l(z)}) corresponds to a case in which
the RH problem can be solved exactly.

It is possible within  the RH approach to handle corrections to
the asymptotic solution to the non-communting RH problem
we have been using in order  to allow us to compute
the response of systems in the intermediate 
regime (where one is interested in the response at frequencies
comparable to those introduced by the perturbation). 
The RH problem lends itself naturally to a type of perturbative
analysis.  The corrections to the
approximate solution valid for long times, (\ref{eq:Y_region1+4}) and 
(\ref{eq:Y_large_t}), can be described by 
multiplying the approximate solution by a function which is
analytic except across
the additional vertical cuts introduced to simplify
the original problem. This function can be specified by a Cauchy 
integral around the cut. Preliminary work in this direction has been
attempted in \cite{Braunecker03_unpublished}.

Finally, the RH method should generalize to non-zero temperatures.
As was observed in \cite{Anderson69}, the singular integral equation
appearing at finite temperatures in a related problem can be solved
analytically. Also, the analytic treatment of the finite-temperature
Fermi-edge singularity in \cite{Ohtaka-Tanabe90,Cobden-Muzykantskii95}
again suggests that the RH approach will generalize successfully to
finite temperatures.

\appendix
\section{Bound States}
If the perturbing potential generates a bound state(s), then
(\ref{eq:FT(S)}) is no longer correct. In the case
where the potential (and hence
$S(t)$) simply switches between its unperturbed value
and a new but time-independent value at $t=0$ and
back again at $t=t_f$, we can correct $\sigma$ by including
the effect of the bound state explicitly. The treatment follows 
closely  that of  \cite{CN71}, although only the case of a separable
potential was treated there.
We write
\begin{equation}
\sigma = \tilde{\sigma} + e^{iH_0 t}|b\rangle e^{-iE_bt} \langle b| . 
\label{eq:sigma_with_b}
\end{equation}
Here $|b\rangle $ is the bound state wavefunction, while
$\tilde{\sigma}$ describes the scattering of the states
within the continuum, and is given by the Fourier
transform of the scattering matrix $S(t)$ (\ref{eq:FT(S)})
as before. 
($H_0$ is the matrix of $\hH_0$ taken between  single-particle
basis states.)

For the case of the function 
$G(t_f) = \mbox{det}| 1 - f + f \sigma |$
(see \ref{eq:core-hole}) we have
\begin{eqnarray}
G(t_f)  & = &  \tilde{G}(t_f)
\mbox{det}\left| 1 + A |b\rangle  \langle b| \right|
\nonumber \\
& = & \tilde{G}(t_f)\left(1 +  A_B \right),
\label{eq:G+bound}
\end{eqnarray}
where $A_B=\langle b| A |b \rangle$ with
\begin{equation}
A =   (1 - f + f \tilde{\sigma})^{-1} f
e^{iH_0 t}e^{-iE_bt}  ,
\label{eq:A_hat}
\end{equation}
and where $\tilde{G}(t_f) =\mbox{det}| 1 - f + f \tilde{\sigma} |$.
We write the bound state as an expansion over the 
basis vectors
\begin{equation}
|b \rangle  =  \sum_\epsilon \bu_\epsilon 
\cdot \ba^\dagger_\epsilon | \rangle.
\label{eq:bound_state}
\end{equation}
For long times $t_f$ the response is dominated by
states within $1/t_f$ of the Fermi energy and it is 
a reasonable approximation to
neglect the energy dependence of the coefficients
$\bu_\epsilon$.
After switching to the time-representation, 
and using (\ref{eq:YfY}) with $\tilde{\sigma}$ in place
of $\sigma$,  we
obtain ($\nu_l$ is the density of states in channel $l$)
\begin{eqnarray}
\langle b| A | b \rangle &  = &  e^{-i E_b t_f} 
\sum_{l l'} \int d\epsilon d\epsilon'
\int dt_1 dt_2 u^*_{l'} u_{l} \times  \label{eq:bAb_inter}
 \\ 
&  &   \hspace{1em} \sqrt{\nu_l \nu_l'}e^{i\epsilon't_1} 
  \left[ Y_+ f Y_-^{-1} 
\right]_{lt_1,l't_2} e^{i\epsilon(t_f-t_2)} .
\nonumber
\end{eqnarray}
Integrating over energies and times gives
\begin{equation}
A_B \sim e^{-i E_b t_f}
u_l \sqrt{\nu_l} \left[ Y_+(0) \frac{1}{-it_f} 
Y_-^{-1}(t_f) \right]_{ll'} 
u^*_{l'} \sqrt{\nu_{l'}}
\label{eq:bAb_final}
\end{equation}
Provided $S(t)$ commutes with itself at all times
between 0 and $t_f$, 
$Y$ is given by (\ref{eq:solution_commuting_S}).
For the single-channel case with $S=e^{2i\tilde{\delta}}$,
we obtain
\begin{equation}
\langle b| A | b \rangle \sim
\frac{\nu}{it_f} \frac{1}{(i\xi_0 t_f)^{2\tilde{\delta}/\pi} } .
\label{eq:bAb_single_channel}
\end{equation}
Here we introduce the quantity $\tilde{\delta}$ which
is the phase shift modulo $\pi$ and takes values on
the interval $[-\pi/2 , \pi/2]$. Normally
the phase shift $\delta$ is defined with a jump
of $\pi$ at a bound state thereby ensuring compliance
with the Friedel sum rule \cite{Friedel52}. However,
when writing the scattering matrix as in (\ref{eq:sigma_with_b}),
the contribution from the bound state to the scattering matrix
is explicitly included in the second term on the right hand side and
is not in the scattering matrix $S$.
At the bottom of the band, the value of 
the phase shift which enters
the threshold shift is clearly $\tilde{\delta}$ as emphasized
in \cite{CN71}.

Although the calculation is longer,
the function $F(t_f)$ can be obtained in a similar
manner   by replacing $\sigma$ in (\ref{eq:det_open_line})
by the form (\ref{eq:sigma_with_b}). One needs only to
keep track of terms up to first order in $e^{-iE_b t_f}$.
(Higher order terms must give zero as they correspond to
double or higher occupancy of the bound state. They can
be seen to make no contribution by subsituting
the formula (\ref{eq:sigma_with_b}) in (\ref{eq:new_one_body}).) 
As for the case of the function $G(t_f)$ considered above,
we neglect the energy dependence of $\bW_\epsilon$ and $\bu_\epsilon$
(see \ref{eq:xray_field} and \ref{eq:bound_state}).
We define
\begin{eqnarray}
C_b & = & (\bW^* \cdot \bu)(\bu^* \cdot \bW) e^{-iE_b t_f} 
\label{eq:C_bound} \\
\tilde{O} & = &  \left( 1 - f + f \tilde{\sigma} \right)^{-1} 
f. 
\label{eq:X_hat}
\end{eqnarray}
Here $\tilde{O}$ is just the scattering state contribution
to $O$ (see \ref{eq:operator_O}):
\begin{equation}
O \simeq \tilde{O} - \tilde{O} e^{iH_0 t_f} |b\rangle e^{-iE_b t_f} 
\langle b|  \tilde{O} .
\label{ref:operator_O_b}
\end{equation}
We obtain (from \ref{eq:F(t_f)_formula})
\begin{eqnarray}
F(t_f) & = & G(t_f) \left[ C - \langle g | O | h \rangle
\right] \nonumber \\
& \simeq & \tilde{G}(t_f) C_b - \tilde{G}(t_f) \left[ 1 + A_B \right]
\langle g | O | h \rangle .
\label{eq:F_B_inter}
\end{eqnarray}
Retaining the dominant terms and ignoring the possibility that
there is an unexpected cancellation between terms proportional
to $e^{-iE_b t_f}$, 
\begin{equation}
F(t_f) = \tilde{G}(t_f) \tilde{L}(t_f)  + a C_b \tilde{G}(t_f) 
\label{eq:open_line+bound}
\end{equation}
where $a \sim 1$ is some constant
and $\tilde{L}(t_f)$ is the scattering state contribution
to the open-line function. For the single channel case with
$S=e^{i2\delta}$, we again assume that the exponent in $\tilde{G}(t_f)$
is $\tilde{\delta} = \delta - \pi$ and obtain 
$F(t_f) \sim F_b(t_f) + F_0(t_f)$ with
\begin{equation}
F_b(t_f) \sim  e^{-iE_bt_f} \frac{1}{(i\xi_0 t_f)^{(\tilde{\delta}/\pi)^2}},
\quad
F_0(t_f) \sim\frac{1}{(i\xi_0 t_f)^{(\tilde{\delta}/\pi-1)^2}}.
\label{eq:open_line+bound_single_channel}
\end{equation}




\section{Computing Spectral Functions}
\label{sec:spectrum}
Given $G(t_f)$ or $F(t_f)$ we would like to compute the
corresponding spectral functions given by a
Fourier integral over $t_f$.
Assume that scattering matrix, $S^e$,  has diagonal elements
$\sqrt{R}e^{i2\alpha_{1,2}}$.
Using the complex cutoff $\zeta V$ (normally $\zeta = i$),
we have from (\ref{eq:Vneq0})
\begin{equation}
\log{G(t_f,V)}   =  -i(E_0-\Delta(V)) t_f - \beta_G \log{(\zeta Vt_f)} + D.
\end{equation}
The exponent $\beta_G = x_+^2 + x_-^2$, where 
\begin{equation}
x_+ = \frac{\log{S^e_{11}}}{2\pi i} =   
 \frac{\alpha_1}{\pi} - i \frac{\log {R}}{4 \pi}  
\label{eq:x+}
\end{equation}
and
\begin{equation}
x_- = \frac{ \log{ \left( 1/S^{e*}_{22} \right) } }{2\pi i} =
\frac{\alpha_2}{\pi} + i \frac{\log {R}}{4 \pi}  .
\label{eq:x-}
\end{equation}
The modified threshold shift is given by (\ref{eq:Delta(V)})
\begin{equation}
\Delta(V) =  
 \left ( \Delta(0) +  V
\frac{\alpha_1 -  \left(\log{S^e}\right)_{11}}{ \pi} \right) 
- i V \left( \frac{\log {R}}{4 \pi} \right)
\label{eq:Delta2(V)} 
\end{equation} 
The real part of $\Delta(V)$  fixes the threshold. We 
will absorb this into the definition
of frequency when computing Fourier transforms.

We write
$\beta_G = \beta_{G1} + i \beta_{G2}$ with
\begin{equation}
\beta_{G1} = \left( \frac{\alpha_1}{\pi} \right)^2 +
\left( \frac{\alpha_2}{\pi} \right)^2 -
\frac{1}{2} \left(\frac{\log{R} }{2 \pi} \right)^2 
\label{eq:betaG1}
\end{equation}
and
\begin{equation}
\beta_{G2} = -  \frac{(\alpha_1 - \alpha_2)}{\pi} \frac{\log R}{2 \pi} .
\label{beta_G2}
\end{equation}
For the function $F(t_f)$, the exponent becomes
$\beta_F = (x_- - 1)^2 + x_+^2$ or
$\beta_F = x_-^2 + (x_+ -1)^2$  
depending on whether the electron is added to the electrode with lower
or higher chemical potential. This gives 
$\beta_F = \beta_G - 2x_\pm + 1$ and
\begin{equation}
\beta_{F1} = \beta_{G1} - \frac{2\alpha_{1,2}}{\pi} + 1 
\quad \mbox{and} \quad
\beta_{F2} = \beta_{G2} \pm \frac{\log R}{2 \pi} .
\label{beta_F}
\end{equation}

Introducing 
\begin{equation}
\omega_2 = - \log{R}/4\pi, 
\label{omega_2}
\end{equation}
the spectral functions of $G$ or $F$ are proportional
to the real part of the Fourier integral,
$\chi_{F,G}({\epsilon})$, where:
\begin{equation}
\chi(\epsilon)  = 
 \int_0^\infty dt_f 
\left( \zeta Vt_f \right)^{-\beta} 
e^{ \left(i\epsilon    -  \omega_2 V  \right)t_f }  ,
\label{eq:chi_integral}
\end{equation}
with $\beta = \beta_F$ for $\chi_F$ and $\beta_G$ for $\chi_G$.
Here the lower limit of the integral is taken to be 0, which is
only valid when $\beta_1 < 1$. When $\beta_1 > 1$,
contributions from the lower limit of the integral 
dominate and the
response is dominated by high frequency contributions which
are not changed from the equilibrium case. These are not described
by the formula (\ref{eq:Vneq0}) and depend on details relating
to the band edge. 
If the phase shifts $x_\pm$ are small, which can be the case for 
the spectral function of $F$  (or  $\tilde{G}A_B$ in the presence 
of a bound state, see \ref{eq:G+bound} and \ref{eq:bAb_single_channel}),
then $\beta_1$ will be close to 1. In this case
$\chi(\epsilon)$ given in (\ref{eq:chi_integral})
contains a significant contribution from times
$t_f < 1/V$ for which our asymptotic solution 
for $F(t_f)$ (or $\tilde{G}A_B$) is incorrect. We can correct
for this by noting that when $1-\beta \ll 1$ the contribution from
times with $Vt_f < 1$ gives just a constant offset which can be
subtracted from $\chi$. To see this, 
we expand the exponential term
$e^{(i\epsilon    -  \omega_2 V )t } $ in the integrand and integrate
term by term from $Vt_f=0$ to $Vt_f=1$. The first term in the expansion
is independent of $\epsilon$ and much larger than subsequent terms
provided $(\epsilon/V) \ll 1/|1 - \beta|$. In practice we subtract
from the real part of $\chi$ its value at $\omega \simeq - V$.
(When $1-\beta$ is not small
the contribution from the times $Vt_f < 1$ to the real part of 
$\chi$ is negligible anyway.)

Eq (\ref{eq:chi_integral}) is in the form of a standard integral and
(see 8.312.2 in \cite{Gradshteyn})
is given by:
\begin{equation}
\chi(\omega_1) 
 =    
(i\zeta)^{-\beta} \frac{i}{V} 
\left(\frac{1}{\omega_1 + i\omega_2}\right)^{1-\beta}
\Gamma(1-\beta) .
\label{eq:chiE}
\end{equation}
If we define $\Omega = |\omega_1 + i \omega_2|e^{i \phi_\Omega}$ and
write:
\begin{equation}
 \Gamma(1-\beta) = |\Gamma(1-\beta)| e^{i\phi_\Gamma} 
\quad \mbox{and} \quad
 i \zeta = e^{i\phi_\zeta},
\label{Omega}
\end{equation}
then
\begin{eqnarray}
\chi(\omega_1)  &  = &  e^{-i\beta_1 \phi_\zeta + \beta_2 \phi_\zeta} 
\frac{i}{V} \frac{e^{i\beta_2 \log{\Omega}}}{\Omega^{1-\beta_1}} \times
\nonumber \\
& & \quad
 e^{i(\beta_1-1) \phi_\Omega -\beta_2 \phi_\Omega} 
 |\Gamma(1-\beta)| e^{i \phi_\Gamma} .
\label{eq:chi(omega1)}
 \end{eqnarray}
The real part of $\chi(\omega_1)$ can then be written \cite{note_on_freqs}:
\begin{eqnarray}
\mbox{Re} \chi(\omega_1) &  = & 
\frac{|\Gamma(1-\beta)|}{V} \frac{1}{\Omega^{1-\beta_1}}
e^{-\beta_2(\phi_\Omega - \phi_\zeta)} 
\label{eq:spectrum}
\\
& & \qquad \sin{\left[ \beta_1(\phi_\zeta - \phi_\Omega ) 
+ (\phi_\Omega - \phi_\Gamma) - \beta_2 \log{\Omega} \right ]}.
\nonumber
\end{eqnarray}
For both functions $F$ and $G$, the cutoff parameter  
$\zeta =  i$, so $\phi_\zeta = \pi$. 





\bibliographystyle{unsrt}
\bibliography{rh}

\begin{thebibliography}{10}
\bibitem{Note_on_sigma}
The definition of $\sigma$ in (\ref{eq:motion}) is the Hermitian conjugate of
  that used in \cite{MA03,MdAB03} to bring the definition into line with the
  more usual convention.
\bibitem{lambda}
It is not essential to introduce the $\lambda$ dependence of $\hat{R}$ exactly
  as in (\ref{eq:R(lambda)}). Any integration of the type $ \log{\hat{A}} =
  \int_{\lambda_1}^{\lambda_2}\hat{A}^{-1} \frac{d\hat{A}}{d\lambda}, $ should
  work provided the limits $\lambda_1$ and $\lambda_2$ are sensibly chosen.
  Here $\lambda$ could be a parameter already involved in the definition of $R$
  or it could be introduced as in (\ref{eq:R(lambda)}).
\bibitem{note_on_freqs}
To compare the response at different voltages requires the use the same scale
  for $\chi$ and would require the inclusion of the constant $D$ in
  (\ref{eq:D}) which depends on $V$. To compare to cases with $\epsilon > V$
  would require matching the long and short time responses. This can be done
  approximately by matching the equilibrium form for the $G(t_f)$ or $F(t_f)$
  valid for short times to the form valid at long times.
\bibitem{note_on_zf}
The factor $z_f = e^{-iVt_f}$ in (\ref{eq:L_n}) measures the energy to move
  from one chemical potential to the other, but its effect on thresholds will
  depend on how the bias voltage is applied. Here we have assumed that $l=2$
  electrode is earthed and the potential on the $l=1$ electrode is varied.
\bibitem{note_on_shot}
In deriving (\ref{eq:shot_final}), we have used the fact that \begin{equation}
  \int dt Y_+ fY_+^{-1} L = \int dt f Y_+Y_+^{-1} L . \label{eq:YfY=fYY}
  \end{equation} This follows from (\ref{eq:Y->1}).
\end{thebibliography}

\end{document}